\documentclass{article}

\usepackage{PRIMEarxiv}

\usepackage[utf8]{inputenc} 
\usepackage[T1]{fontenc}    
\usepackage{hyperref}       
\usepackage{url}            
\usepackage{booktabs}       
\usepackage{amsfonts}       
\usepackage{nicefrac}       
\usepackage{microtype}      
\usepackage{lipsum}
\usepackage{adjustbox}
\usepackage{fancyhdr}       
\usepackage{graphicx}       
\graphicspath{{media/}}     
\usepackage{siunitx}
\usepackage{booktabs}
\usepackage[caption=false,labelformat=simple]{subfig}
\usepackage{flafter} 
\usepackage{placeins}
\usepackage{authblk}
\usepackage{amsmath,amssymb}
\usepackage[font=small,labelfont=bf]{caption}
\ifdefined\qty\else
  \ifdefined\NewCommandCopy
    \NewCommandCopy\qty\SI
  \else
    \NewDocumentCommand\qty{O{}mm}{\SI[#1]{#2}{#3}}
  \fi
\fi
\ifdefined\unit\else
  \ifdefined\NewCommandCopy
    \NewCommandCopy\unit\si
  \else
    \NewDocumentCommand\unit{O{}m}{\si[#1]{#2}}
  \fi
\fi

\makeatletter
\LetLtxMacro\@uthorfrom@uthblk\author

\RenewDocumentCommand{\author}{+o+m}{%
  \ifnum0=\value{authors}%
  \def\AB@authors{}%
  \fi
  \@uthorfrom@uthblk[#1]{#2}%
}
\makeatother

\makeatletter
\renewcommand\AB@affilsepx{, \protect\Affilfont}
\makeatother

\title{\boldmath Shower Separation in Five Dimensions for Highly Granular Calorimeters using Machine Learning}

\author[ ]{\begin{Large}\begin{center}The CALICE Collaboration\end{center}\end{Large}}


\author[a]{S.\,Lai,}
\author[a]{J.\,Utehs,}
\author[a]{A.\,Wilhahn,}

\affil[a]{II. Physikalisches Institut, Georg-August-Universit\"at G\"ottingen, Friedrich-Hund-Platz 1, D-37077 G\"ottingen, Germany}

\author[b]{M.C.\,Fouz,}
\affil[b]{CIEMAT, Centro de Investigaciones Energeticas, Medioambientales y Tecnologicas, Madrid, Spain }

\author[c]{O.\,Bach,}
\author[c]{E.\,Brianne,}
\author[c]{A.\,Ebrahimi,} 
\author[c]{K.\,Gadow,}
\author[c]{P.\,G\"{o}ttlicher,}
\author[c,1]{O.\,Hartbrich,\thanks{{now at Oak Ridge National Laboratory, 1 Bethel Valley Road, Oak Ridge, TN 37830, USA}}}
\author[c]{D.\,Heuchel,}
\author[c,2]{A.\,Irles, \thanks{now at Instituto de Física Corpuscular, Parque Científico, Catedrático José Beltrán, 2 | E-46980 Paterna, España}}
\author[c]{K.\,Kr\"{u}ger,}
\author[c,3]{J.\,Kvasnicka \thanks{also at Institute of Physics, The Czech Academy of Sciences}}
\author[c]{ S.\,Lu,}
\author[c]{C.\,Neub\"{u}ser,}
\author[c]{A.\,Provenza,}
\author[c]{M.\,Reinecke,}
\author[c]{F.\,Sefkow,}
\author[c,4]{S.\,Schuwalow,\thanks{deceased}}
\author[c]{M.\,De Silva,}
\author[c]{Y.\,Sudo,}
\author[c]{H.L.\,Tran,}

\affil[c]{DESY, Notkestrasse 85, D-22603 Hamburg, Germany}

\author[d]{L.\,Liu,}
\author[d]{R.\,Masuda,}
\author[d]{T.\,Murata,}
\author[d]{W.\,Ootani,}
\author[d]{T.\, Seino,}
\author[d]{T.\,Takatsu,}
\author[d]{N.\,Tsuji,}

\affil[d]{ICEPP, The University of Tokyo, 7-3-1 Hongo, Bunkyo-ku, Tokyo 113-0033, Japan }

\author[e]{ R.\,P\"oschl,}
\author[e]{ F.\,Richard,}
\author[e]{D.\,Zerwas,}
\affil[e]{ Université Paris-Saclay, CNRS/IN2P3, IJCLab, 91405 Orsay, France}

\author[e]{F.\,Hummer,} 
\author[e]{F.\,Simon,}

\affil[e] {Karlsruhe Institute of Technology, Institute for Data Processing and Electronics, Kaiserstr. 12, D-76131 Karlsruhe, Germany }

\author[f]{V.\,Boudry,}
\author[f]{J-C.\,Brient,}
\author[f]{J.\,Nanni,}
\author[f]{H.\,Videau,}

\affil[f]{Laboratoire Leprince-Ringuet (LLR), CNRS, \'{E}cole polytechnique, Institut Polytechnique de Paris, F-91120 Palaiseau, France}

\author[g]{E.\, Buhmann,}
\author[g]{E.\,Garutti,}
\author[g]{S.\.Huck,}
\author[g]{G.\,Kasieczka,}
\author[g]{S.\,Martens,}
\author[g,5]{J.\,Rolph\thanks{corresponding author}}
\author[g]{J.\, Wellhausen,}

\affil[g]{Univ. Hamburg, Physics Department, Institut f\"ur Experimentalphysik, Luruper Chaussee 149, 22761 Hamburg, Germany}

\author[h]{B.\,Bilki,}
\author[h]{D.\,Northacker,}
\author[h]{Y.\,Onel,}

\affil[h]{University of Iowa, Dept. of Physics and Astronomy, 203 Van Allen Hall, Iowa City, IA 52242-1479, USA }

\author[i]{L.\,Emberger,}
\author[i]{C.\,Graf}

\affil[i] {Max-Planck-Institut f\"ur Physik, F\"ohringer Ring 6, D-80805 Munich, Germany }

\begin{document}

\maketitle

\begin{abstract}
To achieve state-of-the-art jet energy resolution for Particle Flow, sophisticated energy clustering algorithms must be developed that can fully exploit available information to separate energy deposits from charged and neutral particles. Three published neural network-based shower separation models were applied to simulation and experimental data to measure the performance of the highly granular CALICE Analogue Hadronic Calorimeter (AHCAL) technological prototype in distinguishing the energy deposited by a single charged and single neutral hadron for Particle Flow. The performance of models trained using only standard spatial and energy and charged track position information from an event was compared to models trained using timing information available from AHCAL, which is expected to improve sensitivity to shower development and, therefore, aid in clustering. Both simulation and experimental data were used to train and test the models and their performances were compared. The best-performing neural network achieved significantly superior event reconstruction when timing information was utilised in training for the case where the charged hadron had more energy than the neutral one, motivating temporally sensitive calorimeters. All models under test were observed to tend to allocate energy deposited by the more energetic of the two showers to the less energetic one.  Similar shower reconstruction performance was observed for a model trained on simulation and applied to data and a model trained and applied to data. 
\end{abstract}

\section{Introduction}

A challenging final state jet-energy resolution must be achieved to fulfil the requirements for BSM physics searches and Higgs precision measurements at future linear colliders. For example, for ILC operating at centre-of-mass $\sqrt{s}$= 0.5$-$\qty{1}{\tera \electronvolt} where typical di-jet energies for interesting physics processes will be in the range 150–\qty{350}{\giga \electronvolt}, a relative jet energy resolution of \qty{2.7}{\percent} is crucial \cite{particle_flow}. Particle Flow (PF) is a method expected to provide this resolution, which relies upon accurate tracking of charged particles in a jet, sophisticated event reconstruction techniques, and highly granular sampling calorimeters. A prototype of such a detector is the CALICE Analogue Hadronic Calorimeter (AHCAL) \cite{AHCAL}, a highly-granular steel-scintillator sampling calorimeter designed for PF, with $24\times24\times38$ individual readout cells. The AHCAL is notable for its capacity to measure a timestamp for each readout channel.

Accurate energy clustering algorithms must exploit highly granular calorimeters as part of PF for future linear colliders to achieve this challenging jet energy resolution. Pandora Particle Flow Algorithm (PFA) \cite{particle_flow} is an example of a clustering algorithm for resolving the energy deposits of particles. Explicitly, one of the fundamental tasks of a PFA is the clustering of charged and neutral energy deposits using event information. Furthermore, it has been demonstrated in Ref.~\cite{particle_flow} that the main contributing factor to jet energy resolution for jet energies greater than \qty{50}{\giga \electronvolt} using Pandora PFA is the confusion between the energy deposits of particles. 
Thus, it is scientifically important to minimise confusion by improving techniques for clustering energy deposits from hadron showers. Superior pattern recognition of energy deposits in highly granular calorimeters can achieve this.

Machine learning models provide a powerful tool for developing a bespoke shower separation algorithm using event information. Graph neural networks have been found to provide superior shower separation performance than traditional convolutional neural networks for PF, owing to learning a representation of detector geometry, as demonstrated in Ref.~\cite{gravnet}. However, in Ref.~\cite{gravnet}, there are several significant limitations: 

\begin{itemize}
    \item the sensor density used is more than an order of magnitude smaller than for the AHCAL; 
    \item  charged track position and momentum information are not provided to the neural network, which are critical inputs in PFs to determine the position and scale of the energy belonging to the charged showers; 
    \item the influence of timing information has not been assessed for shower separation.
\end{itemize}

More recent studies on machine learning-based PF, such as in Ref.~\cite{otherpaper1} and Ref.~\cite{otherpaper2}, also do not include timing information as part of the reconstruction. This further motivates complementary research to assess the possible improvements in PF clustering using this additional observable.

The present study evaluated the performance of three published neural network models for hadron shower separation in the AHCAL between a simultaneous charged and synthetic neutral hadron shower produced using data synthesis techniques. These models use information from the event, consisting of hits in the AHCAL sensor array, the track position of the charged particle and its momentum, and aim to predict the fraction of its energy that belongs to either the charged or neutral shower for each hit. The algorithms' performance was then tested using simulation and experimental data, using models trained with and without timing information. 

As an important caveat, this study focuses on the effectiveness of clustering energy deposits from a charged and neutral hadron shower with the AHCAL detector. It neither assesses the effectiveness of the track-cluster association required to 'label' individual energy deposits as either charged or neutral nor the effectiveness of determining the number of simultaneous hadron showers in an event. Results should be interpreted with these caveats in mind.

This paper's superscripts $Q$ and $N$ indicate variables associated with charged and neutral hadrons or showers. 
The variables $E$ and $\widehat{E}$ indicate energy measured by the calorimeter and the value reconstructed by the neural networks. For instance, $E^{N}_{\text{sum}}$ and $\widehat{E}^{N}_{\text{sum}}$ denote the total neutral energy measured by the calorimeter and the total value reconstructed by the neural networks.  Additionally, the predicted and true fractions of energy belonging to a shower in a particular hit of the calorimeter are denoted $\widehat{f}_{\mathrm{hit}}$ and $f_{\mathrm{hit}}$, respectively. Lateral distances between showers are presented in Moliere radii for AHCAL, $\rho_{\mathrm{M}} = \qty{24.9}{\milli \meter}$ \cite{Olin}. Unless otherwise specified, 'data' and 'simulation' are taken to mean experimentally obtained and simulated hadron shower events, respectively.
 

\newpage
\section{Methods and Tools}

\label{sec:CALICE_AHCAL}

\subsection{CALICE AHCAL}

\begin{figure}[!h]
\centering
\subfloat[%
 \label{fig:Physics_AHCAL}
]{\includegraphics[width=0.49\linewidth]{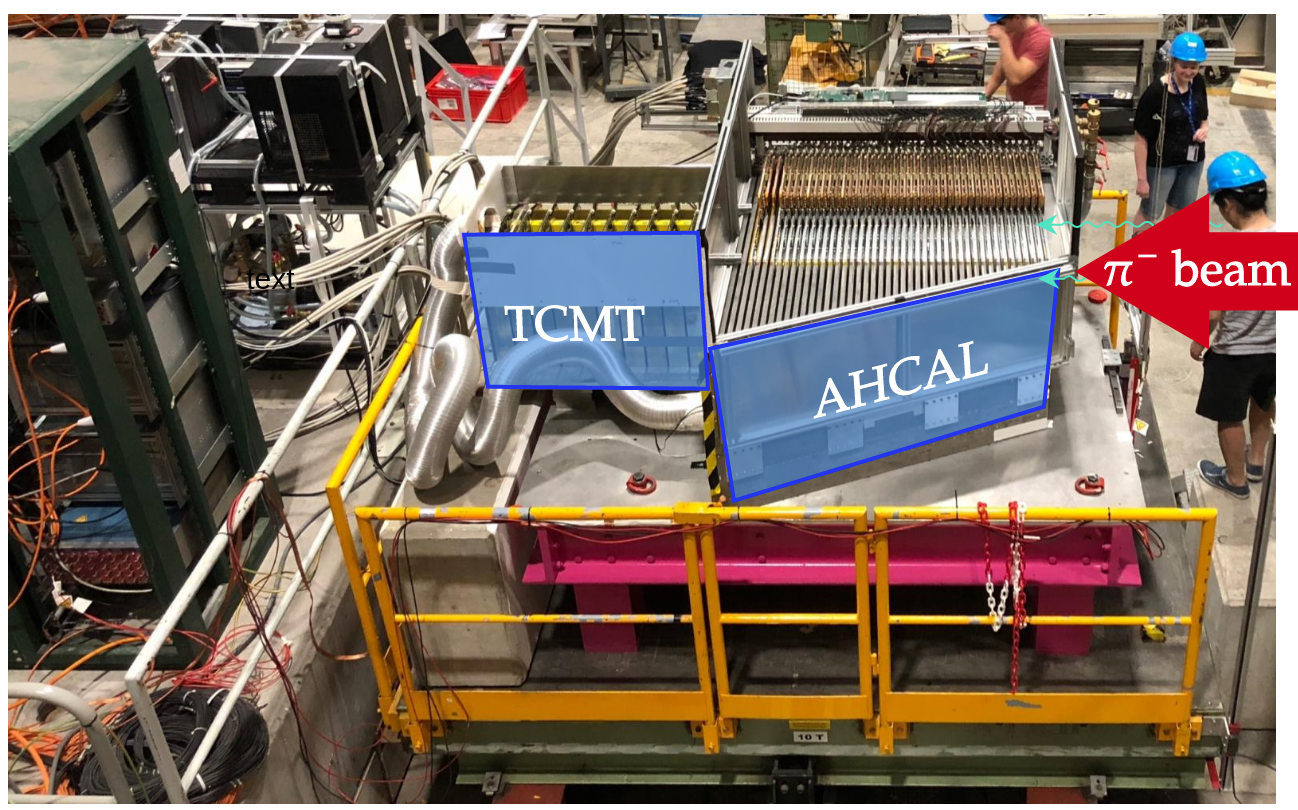}}
\hfill
\subfloat[%
 \label{fig:Physics_AHCAL_IJ}
]{\includegraphics[width=0.44\linewidth]{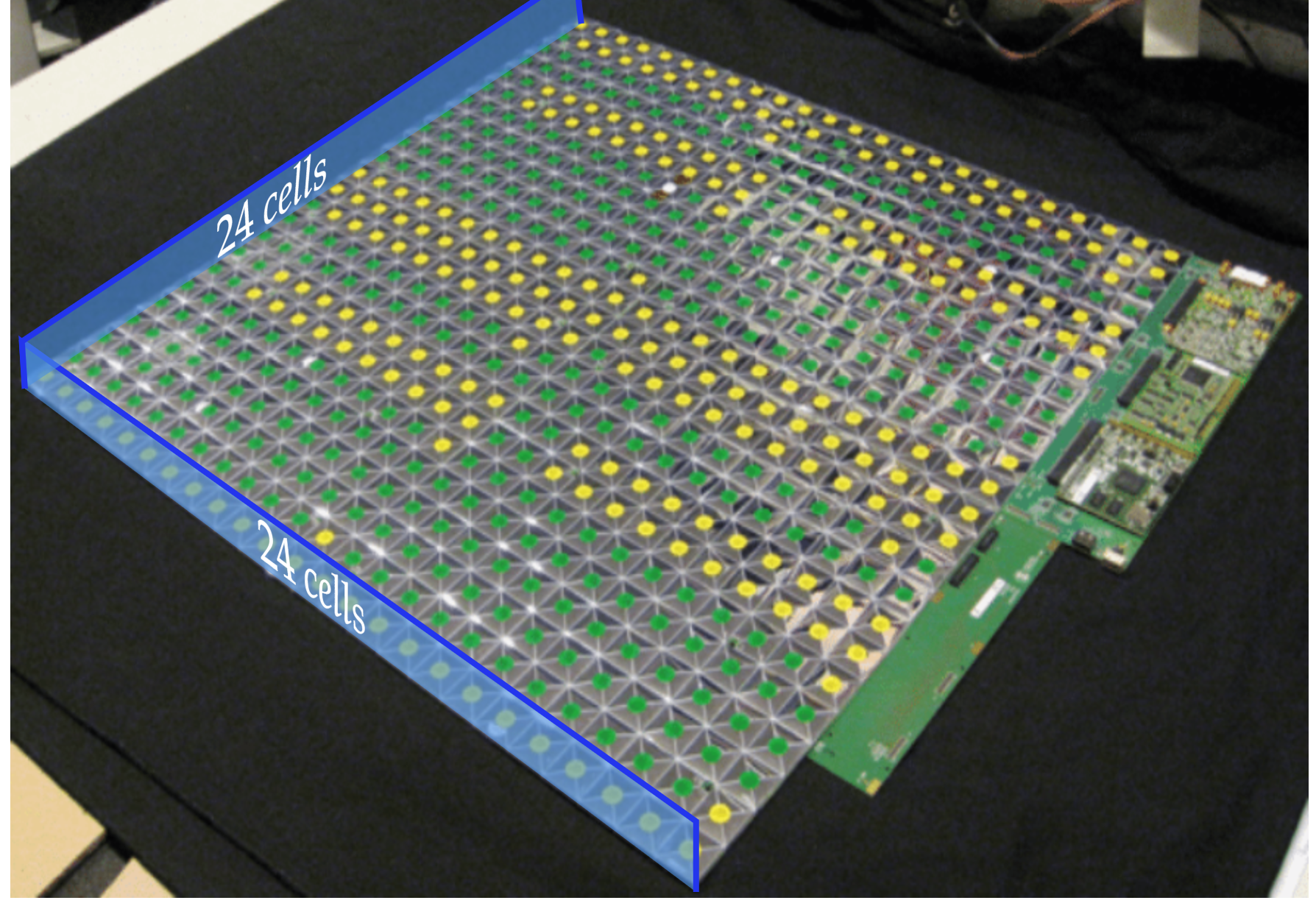}}

\caption{Pictures showing the CALICE AHCAL at testbeam. Fig.~\ref{fig:Physics_AHCAL} shows the detector setup for a testbeam performed in June 2018 at the Super Proton Synchrotron (SPS) at CERN, Geneva \cite{Testbeam}. Fig.~\ref{fig:Physics_AHCAL_IJ} shows one of the 38 layers, with the individual cells of the calorimeter wrapped in foil to improve photon sensitivity.}
\label{fig:AHCAL}
\end{figure}

The CALICE AHCAL is a non-compensating steel-scintillator calorimeter prototype designed for future precision $\mathrm{e}^{+}$-$\mathrm{e}^{-}$ collider experiments. It has a highly granular structure, consisting of $24\times24\times38$ plastic scintillator cells of $30\times30\times \qty{3}{\milli \meter \cubed}$ volume each, read out by individual silicon photomultipliers (SiPMs). These cells indicate the spatial position, magnitude and timestamp of energy deposition with an optimal time resolution of \qty{100}{\pico \second} allowed by the hardware. The detector has a depth of approximately $4.2$ nuclear interaction lengths ($\mathrm{\lambda}_{\mathrm{I}}$). The AHCAL has 38 layers, each with a total depth of \qty{26.1}{\milli \meter}, of which \qty{17.2}{\milli \meter} is steel absorber. The hadronic calorimeter is complemented by a steel-scintillator Tail Catcher/Muon Tracker (TCMT) detector, composed of $320$ extruded scintillator strips of $1000 \times 50 \times \qty{5}{\milli \meter \cubed}$ volume packaged in $16 \times \qty{1}{\meter \squared}$ planes interleaved between steel plates corresponding to an additional depth of \qty{1.1}{\text{\ensuremath{\lambda_{\mathrm{I}}}}} \cite{TCMT}. The TCMT is not used in this analysis. Pictures of the AHCAL calorimeter are shown for reference in Fig.~\ref{fig:AHCAL}. The AHCAL is intended to be part of a complete calorimetric system, including an electromagnetic and hadronic section. The electromagnetic section is not included in this study. This limitation does not prevent relevant studies into algorithms and developments relevant to the hadronic section.

Event information from AHCAL consists of a list of hits (i.e. active cells for which the energy is detected above a noise threshold). The position of a hit indicates the location of an energy deposit in the AHCAL cell matrix ($I_{\mathrm{hit}}$, $J_{\mathrm{hit}}$, $K_{\mathrm{hit}}$). $I_{\mathrm{hit}}$ and $J_{\mathrm{hit}}$ indicate the lateral spatial position of a hit relative to the longitudinal axis of the calorimeter ($I_{\mathrm{hit}}$, $J_{\mathrm{hit}} \in \left[1, 24\right]$ in units of cell index). The longitudinal spatial position (depth in layers) is denoted $K_{\mathrm{hit}}$ ($K_{\mathrm{hit}} \in \left[1, 38\right]$ in units of layer index). The energy ($E_{\mathrm{hit}}$) is measured in units of MIP, the energy deposited by a minimum-ionising particle in a single layer. $E_{\mathrm{hit}}$ is first recorded in Analogue-to-Digital counts and then later converted to the scale of the energy deposited by a minimum ionising particle (MIP) in one cell \cite{AHCALCalib}. $E_{\mathrm{hit}}$ takes a value between a noise threshold of \qty{0.5}{\text{MIP}} and the energy corresponding to the SiPM saturation value. The timestamp ($t_{\mathrm{hit}}$) is defined by the first time at which the electronics signal, proportional to the energy, crosses a given threshold relative to an external trigger. It is then converted to nanoseconds based on a TDC voltage ramp. The ramp's pedestal, maximum value, and time between them are calibrated for each SPIROC2E readout chip of AHCAL \cite{Spiroc} and are used to reconstruct the time value in nanoseconds relative to the trigger. A second deposit later than a few nanoseconds from the first yields no output. Smearing due to electronic noise can result in timestamps less than \qty{0}{\nano \second}. This study considers the ultimate \qty{100}{\pico \second} timing resolution for AHCAL. No charge integration gate length is considered in this study. The calorimeter response is measured as the sum of the individual hits in an event, $E_{\mathrm{sum}} = \sum^{\mathrm{event}} E_{\mathrm{hit}}$. Additionally, the incident position of a charged particle in lateral coordinates is reconstructed using four delay wire chambers (DWC) of $10 \times \qty{10}{\centi \meter \squared}$ size, which is denoted as a vector $\left[ I_{\mathrm{track}}, J_{\mathrm{track}} \right]$ \cite{Testbeam}. The track position of the charged particle entering the calorimeter is a critical input to the shower separation model. Finally, the energy-weighted mean spatial position of the hadron shower in spatial coordinates is defined as a vector called 'centre-of-gravity' ($\mathrm{CoG} = \left[\mathrm{CoG}_{I}, \mathrm{CoG}_{J}, \mathrm{CoG}_{K}\right]$). The shower starting depth is $K_{S}$. $K_{S}$ is calculated using a algorithm described in Ref~\cite{AHCALKS}. Finally, a hit radius is defined, where $R_{\mathrm{hit}}  = \sqrt{(I_{\mathrm{hit}} - \mathrm{CoG}_{I})^2 + (J_{\mathrm{hit}} - \mathrm{CoG}_{J})^2}$, measured in cell units, which describes the distance of an hit to the shower axis.

\subsection{Neural Network Models}\label{sec:NeutralNetworkModels} 

Three neural network models were implemented to assess shower separation for the AHCAL: \texttt{PointNet} \cite{pointnet}, Dynamic Graph Convolutional Neural Network (\texttt{DGCNN}) \cite{dgcnn}, and \texttt{GravNet} \cite{gravnet}. Only \texttt{GravNet} is designed explicitly for PF shower separation of these networks. These neural networks were chosen because they support 'point clouds', a set of sampled points in Cartesian space, a natural representation of a hadron shower in the AHCAL. Explicitly, each active sensor is defined as a point in Cartesian space, $\left[I_{\mathrm{hit}}, J_{\mathrm{hit}}, K_{\mathrm{hit}}, \log{E_{\mathrm{hit}}}, \operatorname{arcsinh }{t_{\mathrm{hit}}}\right]$, where $\operatorname{arcsinh}{t_{\mathrm{hit}}}$ is optional. The transformations of the active hit energy and hit time are used because the distributions of these variables are highly skewed, which makes them poorly suited to machine learning.  In particular, the inverse hyperbolic sine transformation is used to perform a log-like transformation for time information with the possibility of handling smearing for negative values. 

Details of the models can be found in Appendix Section \ref{sec:SummaryNN} and the provided references. The fundamental differences between the models are as follows. \texttt{PointNet} uses a 'global' approach to clustering, exchanging information without considering the local relationships between the individual hits. By contrast, using a dynamically updated graph, \texttt{DGCNN} and \texttt{GravNet} exploit local energy distributions and the relationships between hits. \texttt{DGCNN} directly constructs the graph from the hits, while \texttt{GravNet} projects the hits to a subspace, using these auxiliary hits to cluster. The advantage of \texttt{PointNet} is that it is computationally faster than \texttt{DGCNN} or \texttt{GravNet}, which require sequential $k$-NN clustering as part of the model design. The advantage of \texttt{DGCNN} and \texttt{GravNet}, which are broadly similar in overall design, is that the model can more readily learn local distributions of energy and the relationships between hits, which is expected to result in superior clustering performance compared to \texttt{PointNet}.

The overall designs of the neural networks were modified from the references to reflect the structure of a Pandora PFA algorithm. Explicitly, the first stage of clustering is performed using the positions of the hits only, $\left[I_{\mathrm{hit}}, J_{\mathrm{hit}}, K_{\mathrm{hit}}\right]$, to build a spatial representation of the shower. Track clustering is then encouraged in the next stage by supplying the charged track position and transformed hit energy and hit time $\left[\log{E_{\mathrm{hit}}}, \operatorname{arcsinh }{t_{\mathrm{hit}}}, I^{Q}_{\mathrm{track}}, J^{Q}_{\mathrm{track}}\right]$. In the final stages of each model, track energy ($E^{Q}_{\mathrm{track}}$) and unmodified hit energy information, $\left[E_{\mathrm{hit}},  E^{Q}_{\mathrm{track}}\right]$, is added to encourage a 'statistical re-clustering', involving the aggregation of the energy of a cluster and associating it to a charged track energy. This step is used in Pandora PFA to improve performance for jet energies $E_{j} > \qty{50}{\giga \electronvolt}$, where it is expected that there will be significant confusion between hadron showers. The maximum, mean, and variance were used as aggregation functions in the networks. In certain cases, additional fully connected layers were added to condense the output where necessary. The final output of each network is two sets of fractions, $f^{Q}_{\mathrm{hit}}$ and $f^{N}_{\mathrm{hit}}$, predicting the fraction of energy belonging to $Q$ and $N$ respectively. The sum of $f^{Q}_{\mathrm{hit}}$ and $f^{N}_{\mathrm{hit}}$ is one (i.e. either the energy belongs to $Q$ or $N$). Each network was designed to have around 2 million weights.

.
 
\subsection{Datasets and Training}\label{sec:Datasets}

\subsubsection{Raw Datasets}

Both simulation and experimental data were used for training and evaluation, respectively. Single $\pi^{-}$ hadron shower events observed with the AHCAL detector were studied. The simulation of the particle showers was achieved using \texttt{Geant4} \cite{Geant4}, with a full detector simulation developed using \texttt{DD4hep} \cite{DD4Hep}. Additional effects, such as digitisation of the analogue signal and reconstruction of the detector variables, were achieved for both simulation and data using \texttt{CALICESoft} \cite{CALICESoft}. Timing information from experimental data is not studied due to comparatively poor timing resolution arising from chip occupancy effects \cite{Lorenz}. Useful insights may nonetheless be obtained using timing information in simulation. A MIP-to-GeV calibration factor of \qty{37.3}{\text{MIP} \per \giga \electronvolt} was used \cite{Olin}. The statistics of the training, validation and test datasets are shown in Table \ref{tab:EventTable}. 

Data and simulation were subject to the following cuts:

\begin{itemize}

    \item events were required to be identified using the standard CALICE particle identification algorithm  \cite{PID} as being a single particle and having less than a \qty{0.5}{\percent} probability of being a muon to exclude non-showering, 'punch-through' pions;
    
    \item the 38$^{\mathrm{th}}$ layer of the AHCAL is ganged and requires special treatment beyond the scope of this paper. Therefore, energy deposits were considered up to the 38$^{\mathrm{th}}$ layer of the calorimeter;
    
    \item events were required to have a correctly reconstructed track position (i.e. a track position with a corresponding position inside the $24\times\qty{24}{\text{ cell}}$ AHCAL front-face).

    \item events were required to have at least 50 hits after the MIP-track cut discussed in the following section. This criterion reduces the influence of partially showering punch-through pions, which may initiate a small cascade and continue through the calorimeter.
\end{itemize}

\begin{table*}[h!]
\caption[Number of Events used for Training And Evaluating Shower Separation]{Number of events used for training shower separation models and performing analysis after all cuts.
An additional sample of \qty{40}{\giga \electronvolt} $K^{0}_{L}$ hadrons simulated under the same conditions as the $\pi^{-}$ hadrons is incuded for analysis. Hyphens indicate 0 events. }
\label{tab:EventTable}
\vspace{0.2in}
\centering
\scriptsize{
\begin{tabular}{lllllrrr}
\toprule

Hadron &     $K^{0}_{L}$ & \multicolumn{5}{l}{$\pi^{-}$} & \\
Type &    Simulation & \multicolumn{3}{l}{June 2018 SPS Testbeam Data} & \multicolumn{3}{l}{Simulation} \\
Purpose &      Analysis &                     Testing &     Training & Validation &    Testing &   Training & Validation \\
Particle Energy [\unit{\giga \electronvolt}] &               &                             &              &            &            &            &                 \\
\midrule
5              &             - &                           - &            - &          - &    40685 &    36966 &     4108 \\  
10             &             - &                       21396 &       171166 &      21396 &    68812 &    62333 &     6926  \\
15             &             - &                           - &            - &          - &    75224 &    67379 &     7487  \\
20             &             - &                       32221 &       257762 &      32220 &    77759 &    70866 &     7874  \\
25             &             - &                           - &            - &          - &    81379 &    73023 &     8114  \\
30             &             - &                           - &            - &          - &    80971 &    74180 &     8243  \\
35             &             - &                           - &            - &          - &    78646 &    75619 &     8403  \\
40             &         78146 &                       34428 &       275424 &      34428 &    77055 &    76142 &     8461  \\
45             &             - &                           - &            - &          - &    73620 &    76994 &     8555  \\
50             &             - &                           - &            - &          - &    86014 &    77430 &     8604  \\
55             &             - &                           - &            - &          - &    63218 &    77881 &     8654  \\
60             &             - &                       44600 &       356799 &      44600 &    72306 &    78550 &     8728  \\
65             &             - &                           - &            - &          - &    82256 &    78779 &     8754 \\
70             &             - &                           - &            - &          - &    59806 &    79042 &     8783  \\
75             &             - &                           - &            - &          - &    49390 &    79417 &     8825  \\
80             &             - &                       37790 &       302315 &      37789 &    69106 &    79713 &     8858  \\
85             &             - &                           - &            - &          - &    70091 &    79848 &     8872 \\
90             &             - &                           - &            - &          - &    80773 &    79918 &     8880  \\
95             &             - &                           - &            - &          - &    62647 &    79614 &     8846  \\
100            &             - &                           - &            - &          - &    54433 &    79292 &     8811  \\
105            &             - &                           - &            - &          - &    53377 &    77300 &     8589 \\
110            &             - &                           - &            - &          - &    52632 &    77853 &     8651 \\
115            &             - &                           - &            - &          - &    58750 &    75091 &     8344  \\
120            &             - &                       34881 &       279044 &      34880 &    56274 &    74835 &     8316 \\
\midrule
Total Events            &         78146 &                      205316 &  1642510 &     205313 &  1625224 &  1788065 &   198686  \\
\bottomrule

\end{tabular}
}

\end{table*}

\subsubsection{Data Synthesis and Datasets}\label{sec:DataSynthesis} 

\begin{figure*}
    \centering

    \subfloat[ \label{fig:MIPCut10GeV}]{\includegraphics[width=0.49\linewidth]{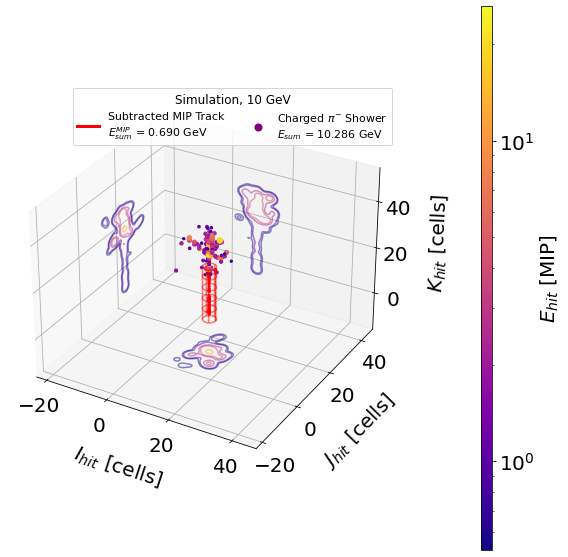}}
    \hfill
    \subfloat[\label{fig:MIPCut80GeV}]{\includegraphics[width=0.49\linewidth]{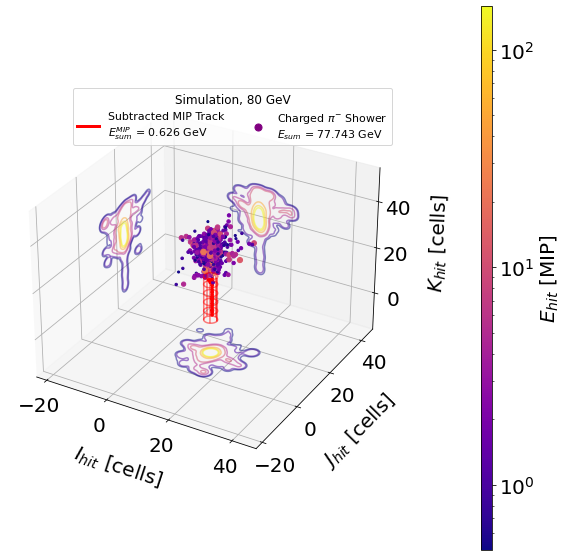}}

    \caption{Event displays of simulated $\pi^{-}$ hadron showers demonstrating the MIP cut applied to two examples from the training sample. Each axis represents the spatial coordinates of the calorimeter, and the purple disks indicate $E_{\mathrm{hit}}$ in a logarithmic scale, both in colour and size. The purple contours on each calorimeter face indicate a smoothed 'energy shadow' of the hadron shower to indicate its profile. Selected MIP-track hits are shown as red circles. The red cylinder indicates the cut region in space. The energy criterion cannot be shown. Figs.~\ref{fig:MIPCut10GeV} and \ref{fig:MIPCut80GeV} show a \qty{10}{\giga \electronvolt} and \qty{80}{\giga \electronvolt} $\pi^{-}$ hadron shower, respectively.}
    \label{fig:MIPCut}
\end{figure*}

Only single $\pi^{-}$ hadron showers from experimental data are available for AHCAL. Therefore, a method is required to produce 'synthetic' showers with two showers, one charged and one neutral. The method must be consistent between the simulation and the data to be compared.

 One of the main differences between charged and neutral hadron showers is that charged particles ionise the detector medium before a shower initiates. A set of highly localised, rectilinear, Landau-distributed energy deposits distributed along the axis of motion of the particle before showering is expected for charged particles. This is called a 'MIP-track'. Therefore, synthetic neutral hadron showers can be produced by applying criteria developed in Ref~\cite{Daniel} to select and remove the MIP-track, referred to as the 'MIP-track cut'. The cut selects hits with $R_{\mathrm{hit}} < \qty{60}{\milli \meter}$, $E_{\mathrm{hit}} < \qty{3}{\text{MIP}}$ and $K\leq  K_{S} -\qty{2}{\text{layers}}$.
 It is noted that the requirement on the hit radius biases the shower shape to symmetric hadron showers. However, only around \qty{3}{\percent} of showers have a track that is further than \qty{60}{\milli \meter} from the centre-of-gravity, which typically indicates poor track reconstruction. Therefore, the effect is minor and necessary. Removing the energy deposits satisfying the cut from the shower results in a synthetic neutral hadron shower. The effect of this cut is shown in Fig.~\ref{fig:MIPCut}.

 This method has clear limitations. For example, it subtracts energy from the shower that would have been part of the shower if initiated by a neutral particle of the same energy. Another limitation is partially showering hadrons, which deposit some energy but continue as MIPs through the calorimeter. Additionally, the likelihood of producing high-energy $\pi^{0}$s is lower for $K^{0}_{L}$ than for $\pi^{\pm}$ due to strangeness conservation, resulting in weaker calorimeter response to $K^{0}_{L}$ compared to $\pi^{\pm}$ of the same energy \cite{wigmans}. However, a study comparing simulated \qty{40}{\giga \electronvolt} $\pi^{-}$ with the MIP-cut applied and simulated \qty{40}{\giga \electronvolt} $K^{0}_{L}$ hadron showers found them similar at the event level, making them acceptable replacements for neutral showers. Further study details are provided in Appendix Section~\ref{sec:ROCTest}.

\begin{figure*}
    \centering

    \subfloat[
      \label{fig:2ShowerComb}
    ]{\includegraphics[width=0.49\linewidth]{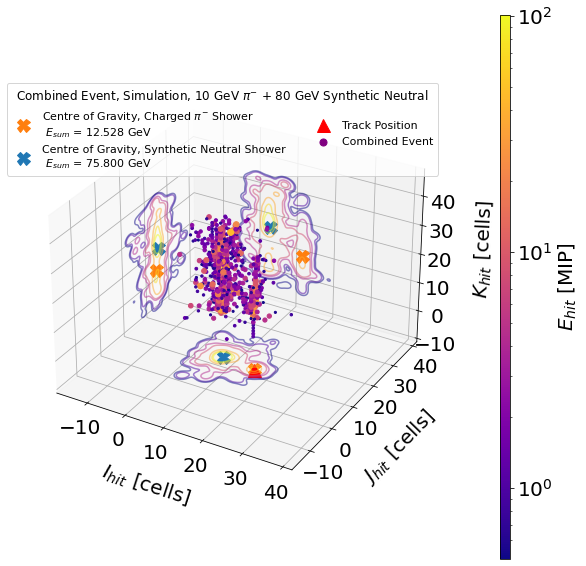}}
    \hfill
    \subfloat[
    \label{fig:2ShowerSep}
    ]{\includegraphics[width=0.49\linewidth]{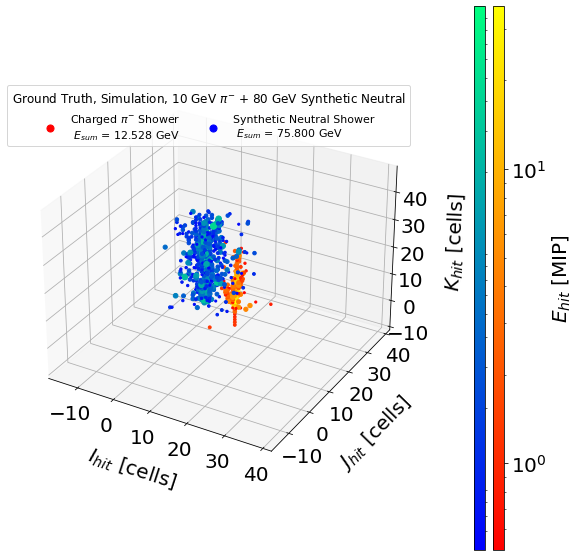}}
   
\caption{Fig.~\ref{fig:2ShowerComb} shows an event display of two overlayed simulated showers, one \qty{10}{\giga \electronvolt} charged and \qty{80}{\giga \electronvolt} synthetic neutral, where track position and centres of gravity of the charged and neutral shower indicated by a red triangle, and an orange and blue cross, respectively. Fig.~\ref{fig:2ShowerSep} shows the same event as Fig.~\ref{fig:2ShowerComb}, with the individual showers identified by colour. The red and blue points indicate the simulated charged $\pi^{-}$ hadron shower and a synthetic neutral hadron shower, respectively. Else, as in Fig.~\ref{fig:MIPCut}.   }
\label{fig:2Shower}
\end{figure*}


A pair of showers are then overlaid by shifting their lateral positions within a circle to create synthetic events. The circle's radius is based on the radial distance from the centre-of-gravity within which 80\% of each shower's energy is contained and depends on the particle energy. Overall, $Q$ and $N$ have a most-probable distance of \qty{5.5}{\ensuremath{\rho_{M}}} apart. For each cell which has deposited energy from both $Q$ and $N$, $E_{\mathrm{hit}}$ and $t_{\mathrm{hit}}$ are recalculated as the sum of the energies of the original showers and the earliest timestamp of the two hits. $f^{Q}_{\mathrm{hit}}$ and $f^{N}_{\mathrm{hit}}$ are calculated as the fraction of energy in the cell belonging to either $Q$ or $N$. The specific details of the method, including how the average inter-shower distance for the data was selected, are discussed in Appendix Section \ref{sec:ShowerAlg}. Appendix Fig.~\ref{fig:Algorithm} provides a flow-chart of the algorithm for reference. The result of the method is a combined shower with fractions of hit energy belonging to each shower, $f^{Q}_{\mathrm{hit}}$ and  $f^{N}_{\mathrm{hit}}$, that are to be reconstructed by the neural networks.

For simulation, $7.2\times10^{5}$ and $8\times10^{4}$ synthetic charged-neutral hadron showers with an average of $1250 \pm 37$ events and $139 \pm 12$ events per particle energy combination were produced for training and validating the neural networks during the training phase, respectively, while $8\times10^{5}$ events were produced to test the models, with an average of $1389 \pm 39$ events per particle energy combination. Each combined sample contained showers purely from the corresponding source samples outlined in Table \ref{tab:EventTable}. The same number of events for training and validation samples were chosen for data. However, a smaller sample of $2\times10^{5}$ events were used for testing than for simulation, owing to a smaller sample of available test events. Also, the average number of events per particle energy for training, validation, and testing in data was $20000 \pm 140$ events, $2222 \pm 40$ events and $5556 \pm 81$ events, respectively. The number of events in data is higher than for simulation due to fewer available energy combinations.

\subsubsection{Training}

For simulation, two independent neural networks based on each model defined in Section \ref{sec:NeutralNetworkModels} were trained with and without timing information. For data, a single neural network with the best performance in the simulation was trained without timing information and tested on data. Another instance of the same model was then trained and tested on data. The networks were developed in \texttt{PyTorch} \cite{PyTorch} and trained using the \texttt{PyTorch}lightning research framework \cite{PyTorchLightning} on an NVidia V100 GPU. The ADAM optimiser was used to improve the convergence rate for ten epochs.  The hyperparameters used for training are shown in Table \ref{tab:Hyperparameters}, selected based on the results of a parameter scan using Optuna hyperparameter optimisation framework \cite{Optuna}. It is noted that the $\gamma$ parameter of \texttt{GravNet} was also varied as a hyperparameter, which was not done in Ref~\cite{gravnet}.

\begin{table*}
    \caption[Table of Hyperparameters for Shower Separation Training]{Hyperparameters used to train the neural network. In this table, $\beta_{1}$ and $\beta_{2}$ are the ADAM momentum parameters, $p_{\mathrm{dropout}}$ is the dropout probability, and $k$ is the number of nearest-neighbours per cluster. Hyphens indicate hyperparameters that do not apply to the model. The parameters were informed by a hyperparameter scan using Optuna \cite{Optuna}.}
    \centering
    \scriptsize{
    \adjustbox{width=\textwidth}{
    \begin{tabular}{rrrrrrr} 
    \toprule
    Parameter & PointNet, no Time & PointNet, + Time & GravNet, no Time & GravNet, + Time & DGCNN, no Time & DGCNN, + Time \\
    \midrule 
    Learning Rate & $2.567 \times 10^{-4}$ & $5.681 \times 10^{-5}$ & $2.012 \times 10^{-4}$ & $5.169 \times 10^{-4}$ & $1.252 \times 10^{-5}$ & $1.660 \times 10^{-4}$ \\
 $p_{\text{dropout}}$ & $0.332$ & $0.259$ & $0.268$ & $0.469$ & $0.167$ & $0.164$ \\
$\gamma$ & - & - & $8.137$ & $12.815$ & - \\
$k$ & - & - & $16$ & $24$ & $15$ &  $18$\\
    $\beta_{1}$  & $0.9$ & $0.9$ & $0.9$ & $0.9$ & $0.9$ & $0.9$\\
    $\beta_{2}$  & $0.99$ & $0.99$ & $0.99$ & $0.99$ & $0.99$ & $0.99$\\
    \bottomrule
\end{tabular}
    }
    }
   
    \label{tab:Hyperparameters}

\end{table*}

 The loss was chosen to be the same as in the study of Ref~\cite{gravnet}. This study applied a square-root energy-weighted mean square loss during training to encourage the models to cluster the most energy-dense parts of the shower correctly. However, to reduce the influence of 'shower-swapping' (i.e. when the neural network correctly separates the two showers but compares them to incorrect permutations during training and evaluation, see Ref~\cite{gravnet} for more details), a permutation-invariant training approach was adopted.

 The loss is shown in Eq.~\ref{eq:ShowerSep_LossFunction}:


\begin{equation}
     \mathcal{L}(\widehat{f}_{\mathrm{hit}};  f_{\mathrm{hit}}, E_{\mathrm{hit}} )  = 
    \sum^{\{Q, N\}}_{i=0} \frac{  \sum_{\text{event}}\sqrt{E_{\mathrm{hit}}\cdot f^{i}_{\mathrm{hit}}} \cdot (\widehat{f}^{i}_{\mathrm{hit}} - f^{i}_{\mathrm{hit}})^{2}}{ \sum_{\text{event}} \sqrt{E_{\mathrm{hit}}\cdot f^{i}_{\mathrm{hit}}} }
    \label{eq:ShowerSep_LossFunction}
\end{equation}

{\setlength{\parindent}{0cm}
where $\{ Q, N \}$ is the set of possible hadron showers, $Q$ and $N$ are charged and neutral hadrons inducing showers in the AHCAL, and $i$ is the index of each set element. 
}

The loss is then found for each possible permutation of outputs from the network by re-ordering the output from each network, which could be either $\{f^{Q}_{\mathrm{hit}}, f^{N}_{\mathrm{hit}} \}$ or $\{f^{N}_{\mathrm{hit}}, f^{Q}_{\mathrm{hit}} \}$. The minimum loss of the of permutations is then taken. This consideration helps reduce the influence of 'shower-swapping' on the network's performance during training.
 
The results must be interpreted with the following biases in mind: 
\begin{itemize}
    \item the study focuses solely on scenarios involving one charged and one neutral hadron shower, without considering an unknown number of showers;

    \item there is ambiguity in the order of the output of each model to overcome the confusion of 'shower-swapping' (i.e. the model does not label each shower as $Q$ or $N$, but instead focuses entirely on clustering the energy deposits);

    \item the distance distribution between incident hadrons is ad-hoc and chosen to train a robust and functional shower separation algorithm. In the ideal case, the inter-particle distance of a jet, which could be obtained from simulation and was not available for this study, would be used instead. 
\end{itemize}

Finally, for the \texttt{PointNet} network only, the $A$ matrix as defined in Appendix Section \ref{sec:SummaryNN} is constrained to be close to an orthogonal matrix by adding a regularisation condition to the loss of Eq.~\ref{eq:ShowerSep_LossFunction} to produce a modified loss function. It is shown in Eq.~\ref{eq:PointNetLoss} \cite{pointnet}:

\begin{equation}
    \mathcal{L}_{\mathrm{reg}}(A) = \left\|I-A A^T\right\|^2 
    \label{eq:PointNetLoss}
\end{equation}

where $I$ is the identity matrix, superscript $T$ indicates the matrix transpose operation.

\section{Results}

In this section, the shower separation capabilities of each neutral network presented in Section \ref{sec:NeutralNetworkModels} are presented and analysed. 


The neutral hadron shower, $N$, is considered the reference shower henceforth. Confusion energy is therefore defined as the predicted minus the true reconstructed calorimeter response:

\begin{equation}
    E^{N}_{\text{confusion}} (\widehat{E}^{N}_{\text{sum}};   E^{N}_{\text{sum}}) = \widehat{E}^{N}_{\text{sum}} - E^{N}_{\text{sum}}
    \label{eq:Confusion}
\end{equation}

\subsection{Reconstruction Quality and Confusion Distribution}

The energy distributions of reconstructed hadron showers are first evaluated for the original shower energy for the three separation models. The reconstruction quality for reconstructed neutral showers is evaluated using the mean and the most probable value (MPV) obtained from the confusion energy distribution, defined in Equation \ref{eq:Confusion}. The MPV is estimated using a kernel density estimate using \texttt{KDEpy} \cite{KDEPy}, with Silverman's binning rule utilised to determine the bandwidth \cite{Silverman}. A spline is then fitted to the estimate, and the MPV is determined by locating the root of the spline with the maximum probability density.

The confusion energy distributions of the implemented shower separation models are assessed, presenting the $\text{RMS}_{90}$ and median absolute deviation (MAD) of each distribution. These statistics are used to compare their resilience to extreme outliers to standard deviation measures. $\text{RMS}_{90}$ is defined as the minimum standard deviation within all possible central \qty{90}{\percent} percentile ranges permitted by the data, and is commonly used in the Particle Flow community to measure calorimeter response spread. The MAD is defined as the median distance of the absolute deviations of the data to the median \cite{MAD} and serves as an additional robust statistic commonly employed in statistical analysis.

\begin{figure*}
    \vspace{-0.5in}
    \centering

    \subfloat[ \label{fig:Lin_PN_nt_sim}]{\includegraphics[width=0.4\linewidth]{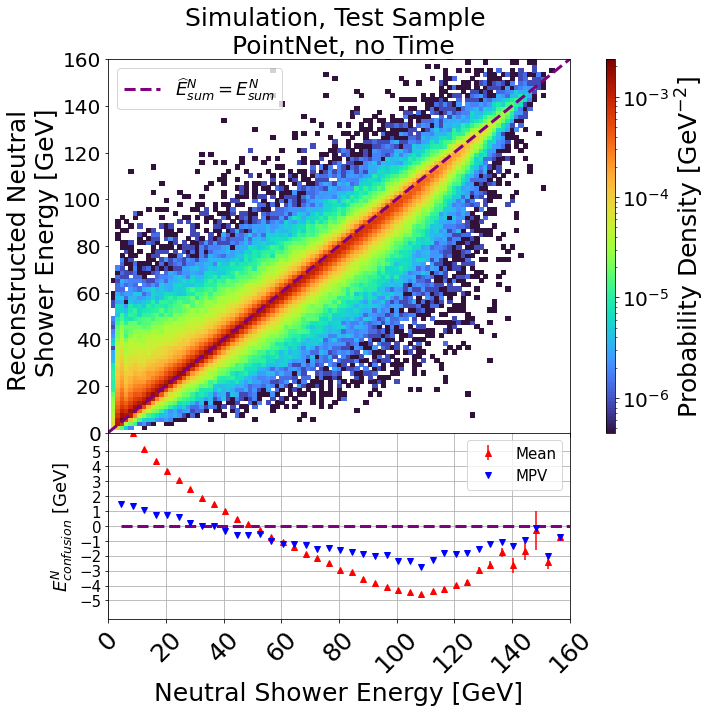}}
    \hfill
    \subfloat[\label{fig:Lin_PN_wt_sim}]{\includegraphics[width=0.4\linewidth]{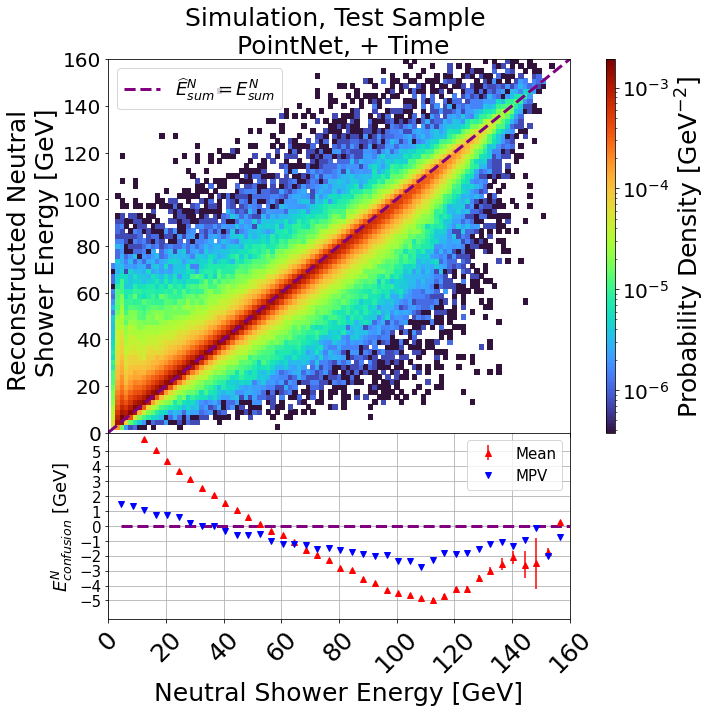}}

    \subfloat[ \label{fig:Lin_DGCNN_nt_sim}]{\includegraphics[width=0.4\linewidth]{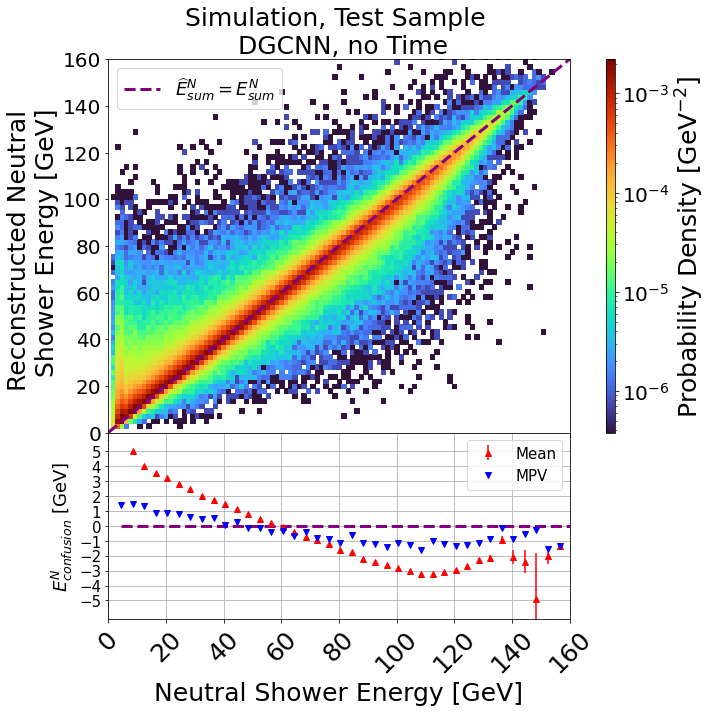}}
    \hfill
    \subfloat[\label{fig:Lin_DGCNN_wt_sim}]{\includegraphics[width=0.4\linewidth]{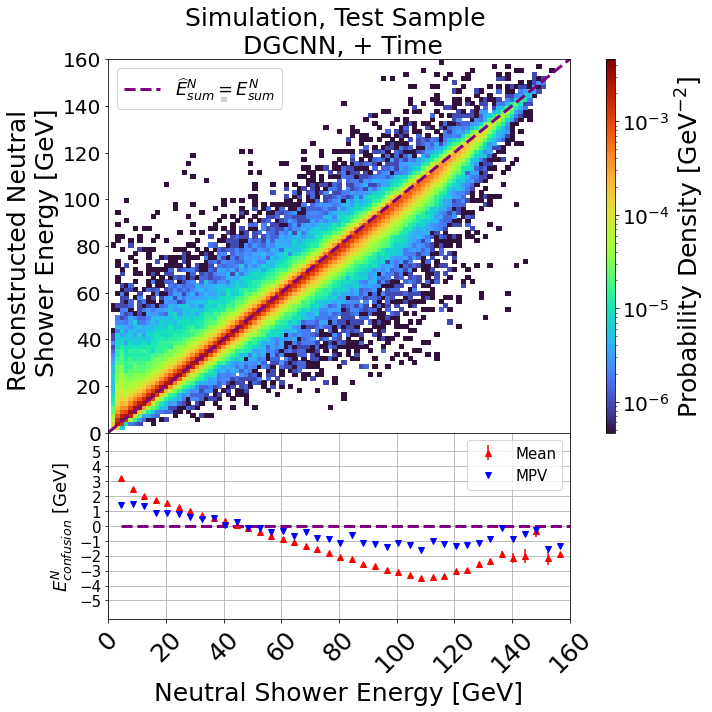}}

    \subfloat[ \label{fig:Lin_GN_nt_sim}]{\includegraphics[width=0.4\linewidth]{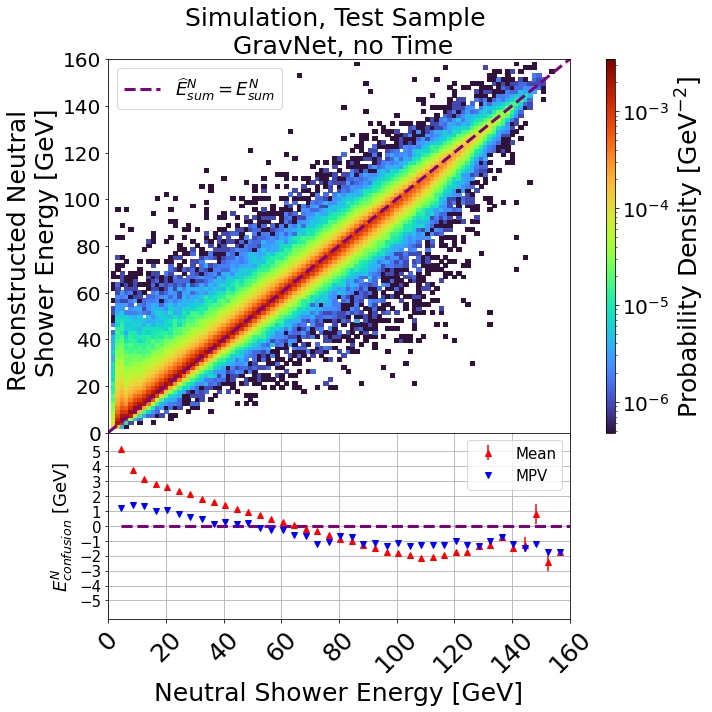}}
    \hfill
    \subfloat[\label{fig:Lin_GN_wt_sim}]{\includegraphics[width=0.4\linewidth]{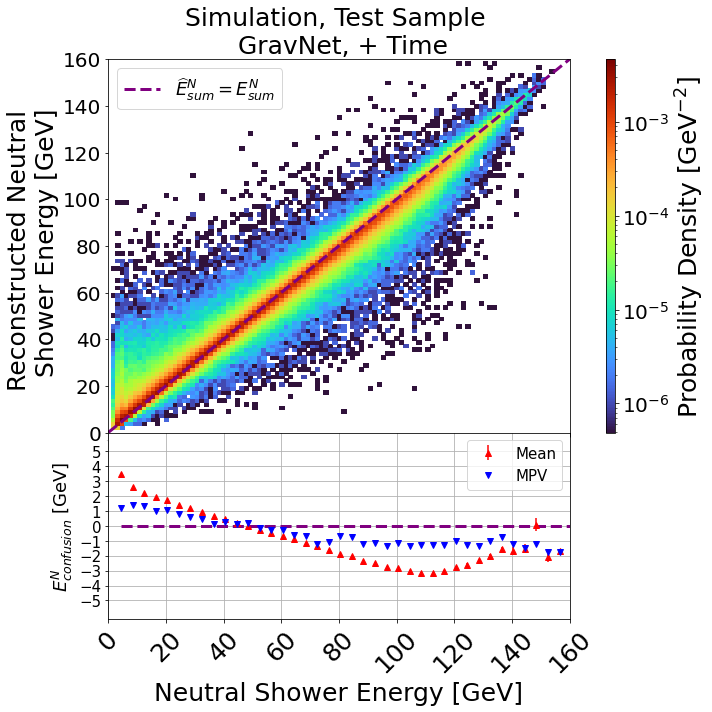}}

    \caption{Figs.~\ref{fig:Lin_PN_nt_sim}, \ref{fig:Lin_DGCNN_nt_sim}, \ref{fig:Lin_GN_nt_sim} (left) and  Figs.~\ref{fig:Lin_PN_wt_sim}, \ref{fig:Lin_DGCNN_wt_sim}, \ref{fig:Lin_GN_wt_sim} (right) show the joint distributions of the predicted and true reconstructed neutral shower response for \texttt{PointNet}, \texttt{DGCNN} and \texttt{GravNet}, without and with timing information, respectively. The colour scale indicates probability density. The purple dashed line indicates perfect reconstruction. The bottom subplot shows the mean and MPV on the $y$-axis at each bin along the $x$-axis.}
    \label{fig:Lin_sim}
\end{figure*}

\paragraph{Simulation} The reconstruction quality for each model applied to the testing dataset is depicted in Figure \ref{fig:Lin_sim}. This figure illustrates that, in general, the models tend to reconstruct the neutral shower with energy levels close to the original. The MPVs of confusion energy, as shown in the subplots, exhibit differences of no more than \qty{1.5}{\giga \electronvolt}. However, a bias is evident from the mean values in the subplot and the green regions in the main figures, indicating a tendency to overestimate the shower energy of neutrals below \qty{60}{\giga \electronvolt} and underestimate it above that value. This bias is observed across all models under the test, evident from the steeper slope of the mean compared to the MPV of each distribution. The discrepancy between the MPV and mean, along with the asymmetry of the green and blue regions around shower energies of \qty{60}{\giga \electronvolt}, indicates skewness in confusion distributions. Comparing the left column with the right column of figures reveals significant improvements in neutral hadron shower reconstruction for \texttt{DGCNN} and \texttt{GravNet} when incorporating \qty{100}{\pico \second} time resolution, as evidenced by the narrowing of the distributions around the purple line. Conversely, \texttt{PointNet} shows no such improvement.


\begin{figure*}
    \centering

    \subfloat[ \label{fig:EConf_PN_sim}]{\includegraphics[width=0.42\linewidth]{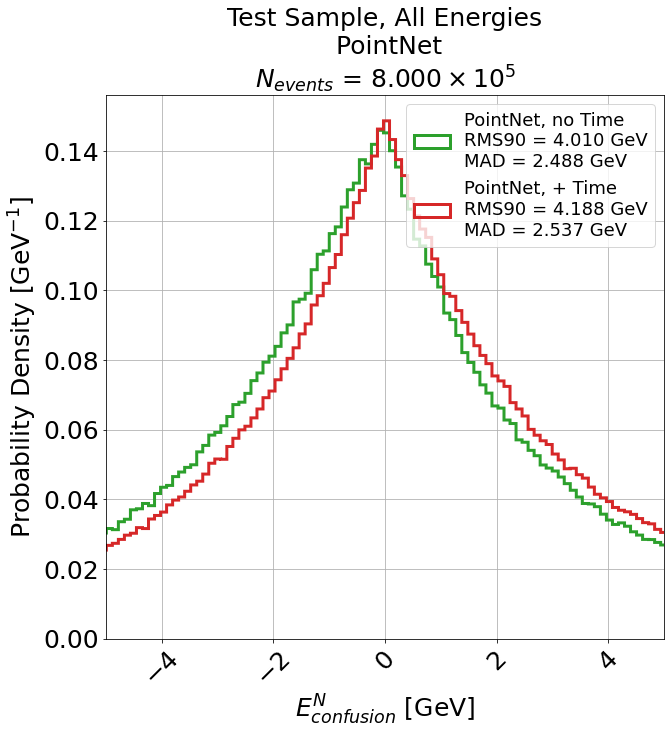}}
    \hfill
    \subfloat[\label{fig:EConf_DGCNN_sim}]{\includegraphics[width=0.42\linewidth]{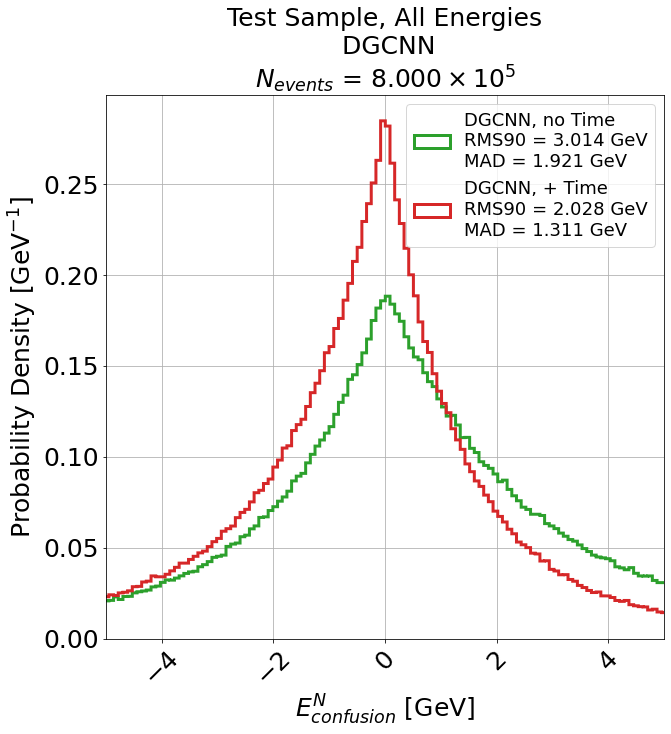}}
    
    \subfloat[ \label{fig:EConf_GN_sim}]{\includegraphics[width=0.42\linewidth]{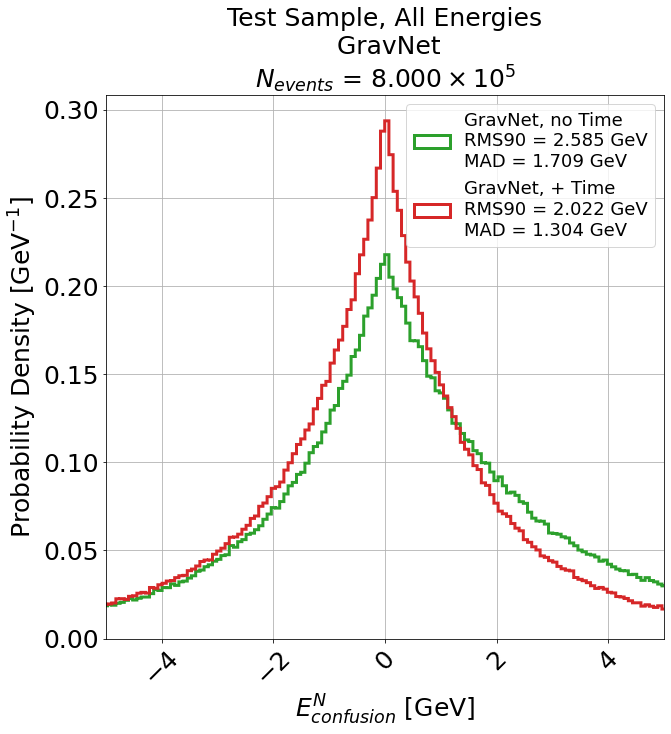}}

    \caption{
    Distributions of neutral confusion energy for each shower separation model under test. The green and red lines indicate the same models, without and with timing information, respectively. $\text{RMS}_{90}$ and MAD are shown in the legend for each model.}
    \label{fig:EConfusion_sim}
\end{figure*}

The confusion energy distributions for each model under test are depicted in Fig.~\ref{fig:EConfusion_sim}. Comparing Figure \ref{fig:EConf_PN_sim} with Figures \ref{fig:EConf_DGCNN_sim}-\ref{fig:EConf_GN_sim} reveals that the \texttt{PointNet} model performs similarly in reconstructing hadron showers, regardless of whether timing information is utilised. In contrast, both the \texttt{DGCNN} and \texttt{GravNet} models exhibit significantly less confusion compared to \texttt{PointNet} when timing information is provided to these models. This improvement in performance could be attributed to the ability of graph neural networks (\texttt{DGCNN} and \texttt{GravNet}) to exploit local geometric structures (i.e. patterns of energy density in space and time), unlike \texttt{PointNet}, which treats energy deposits independently \cite{dgcnn}. Notably, a slight positive skewness in the distribution is observed for \texttt{DGCNN} and \texttt{GravNet} models (i.e. where the MPV of the distribution is displaced from 0), without timing information. The reasons for this are unknown.

Overall, the \texttt{GravNet} model performed the best of the models under test, both with and without timing information. Without timing information, \texttt{GravNet} demonstrated less confusion and comparable performance to \texttt{DGCNN} with timing. Introducing \qty{100}{\pico \second} timing resolution led to a notable \qty{23}{\percent} reduction in the median absolute deviation (MAD) of the confusion distribution for \texttt{GravNet}.  The optimal 'potential scaling' parameter, $\gamma$, increased with timing information inclusion (see Table.~\ref{tab:Hyperparameters}), indicating a more localised influence of individual cells in the shower and suggesting a more sophisticated clustering approach. It is noted that no strong correlations were observed between typical shower observables and the improvement due to timing information. Therefore, further research is needed to grasp the full influence of timing information on clustering and how this is achieved. It is also noted that in a real jet, this improvement due to timing information will likely improve even further due to the delay of charged particles, which bend in a magnetic field and thus travel further to get to the calorimeter.

\begin{figure*}
    \vspace{-0.5in}
    \centering

    \subfloat[ \label{fig:Skew_PN_nt_sim}]{\includegraphics[width=0.4\linewidth]{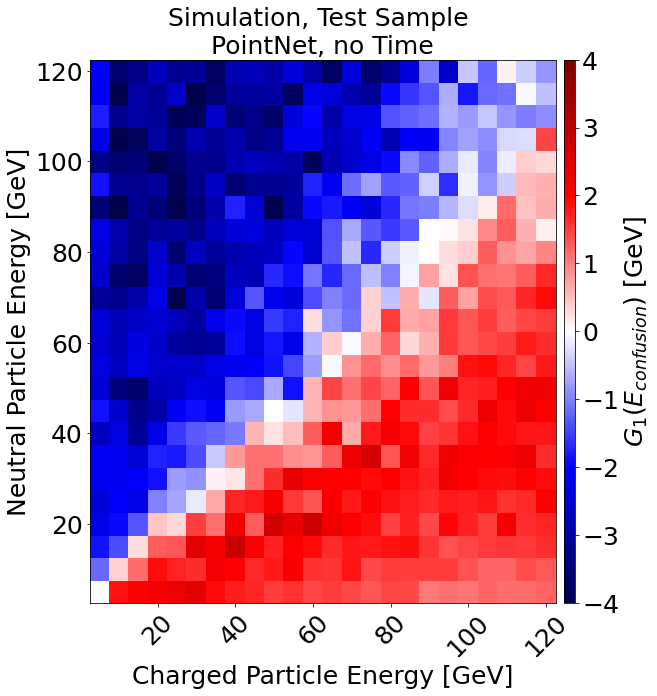}}
    \hfill
    \subfloat[\label{fig:Skew_PN_wt_sim}]{\includegraphics[width=0.4\linewidth]{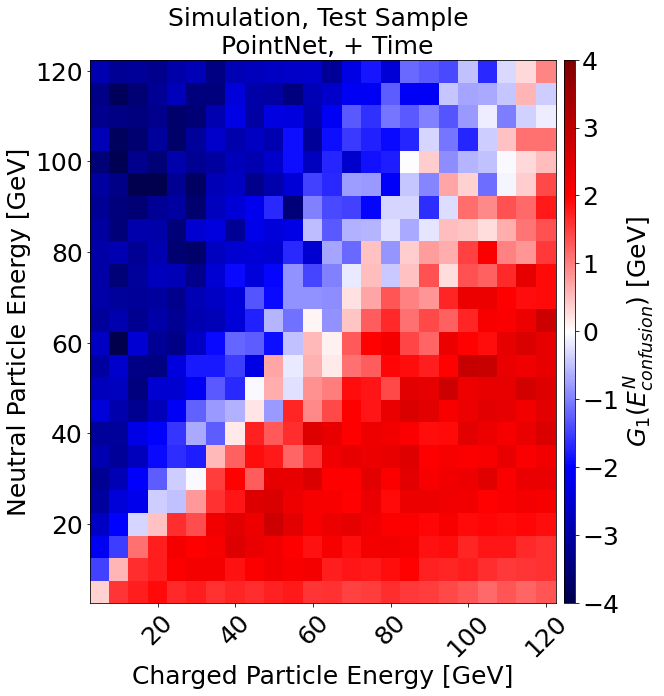}}

    \subfloat[ \label{fig:Skew_DGCNN_nt_sim}]{\includegraphics[width=0.4\linewidth]{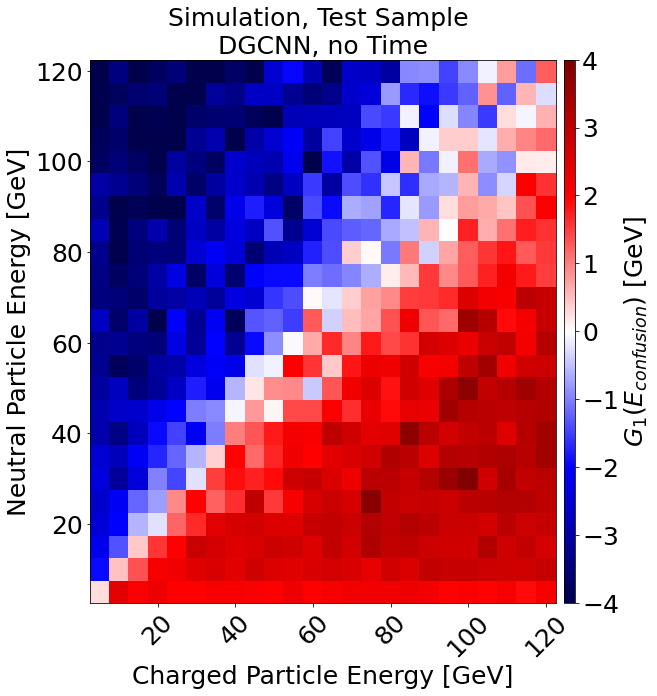}}
    \hfill
    \subfloat[\label{fig:Skew_DGCNN_wt_sim}]{\includegraphics[width=0.4\linewidth]{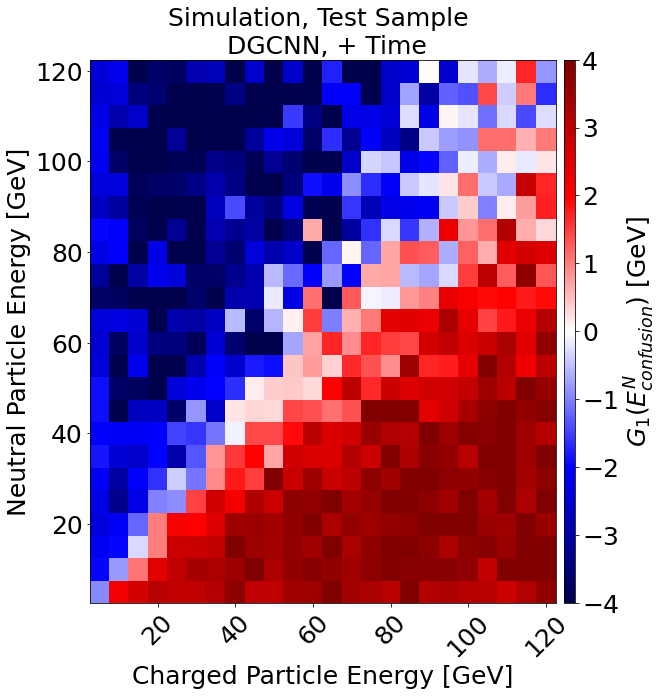}}

    \subfloat[ \label{fig:Skew_GN_nt_sim}]{\includegraphics[width=0.4\linewidth]{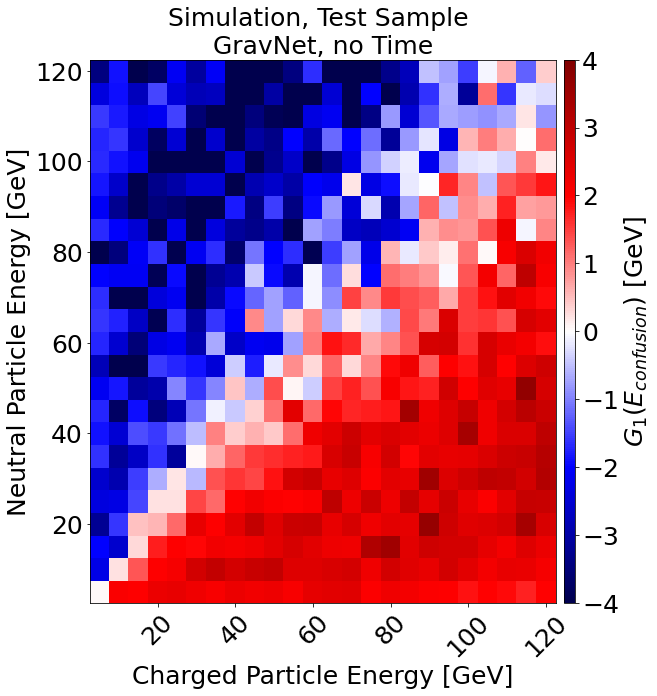}}
    \hfill
    \subfloat[\label{fig:Skew_GN_wt_sim}]{\includegraphics[width=0.4\linewidth]{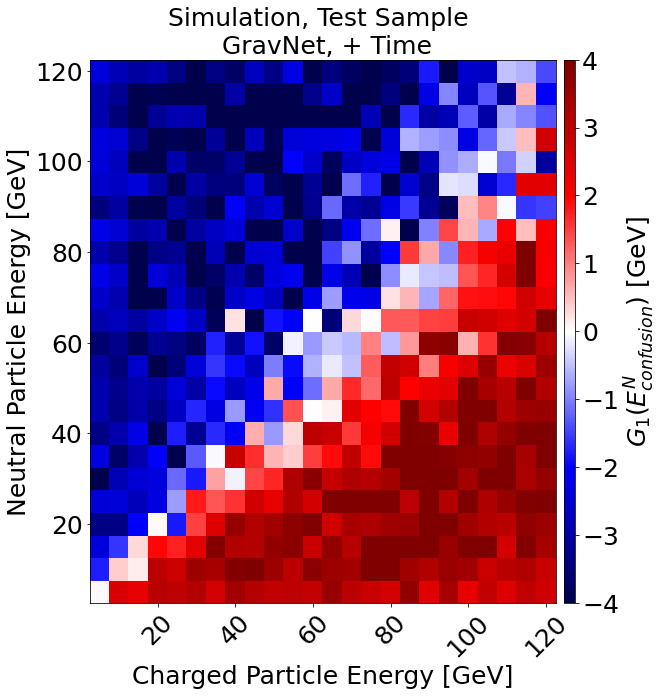}}

    \caption{Skewness as a function of charged and neutral shower energy. The $x$ and $y$-axes indicate the charged and neutral shower energy, respectively. The colour scale indicates skewness, $G_{1}$, where red and blue indicate positive and negative skewness, respectively.}
    \label{fig:Skew_sim}
\end{figure*}

The skewness of confusion energy distributions for each tested model is displayed using the adjusted Fisher-Pearson standardised moment coefficient ($G_{1}$) \cite{Skew}, as depicted in Fig.~\ref{fig:Skew_sim}.

All models exhibit positive skewness when the charged particle energy exceeds the neutral one, implying an overestimation of neutral shower energy. Similarly, underestimation is observed in the opposite case. This pattern suggests that all methods of shower separation lean towards an 'altruistic' approach, where higher-energy showers transfer energy to lower-energy ones. A plausible explanation for this phenomenon is that redistributing energy from the highest-energy hadron shower to lower-energy ones during clustering is an optimal strategy. Notably, Pandora PFA, a clustering algorithm, tends to split true clusters during its initial stage rather than merging energy deposits from multiple particles into a single cluster \cite{particle_flow}. Similar skewness in the confusion distribution has been independently observed in other studies involving Pandora PFA \cite{Daniel}. Further analysis is required to validate these hypotheses beyond the presented studies.

 

\begin{figure*}
    \centering

    \subfloat[ \label{fig:EConf_GN_rightskew_sim}]{\includegraphics[width=0.49\linewidth]{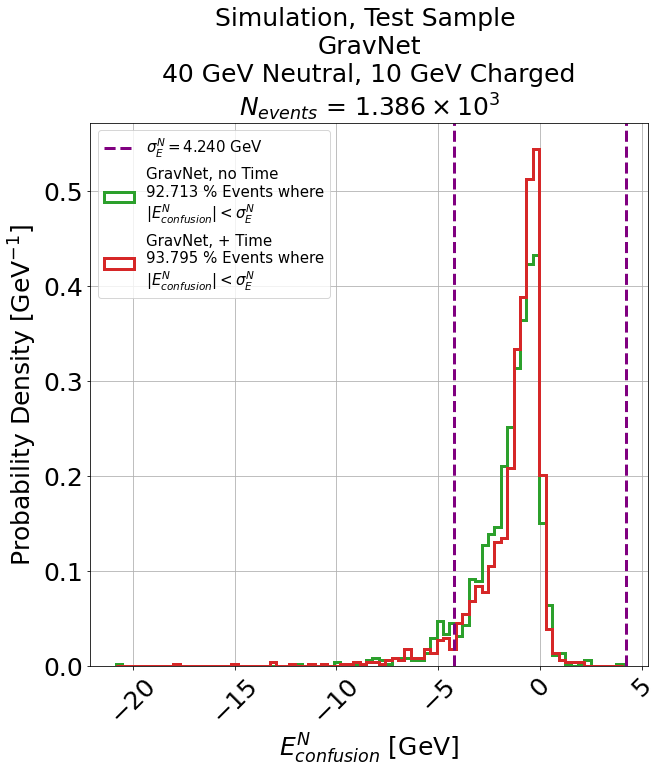}}
    \hfill
    \subfloat[\label{fig:EConf_GN_leftskew_sim}]{\includegraphics[width=0.49\linewidth]{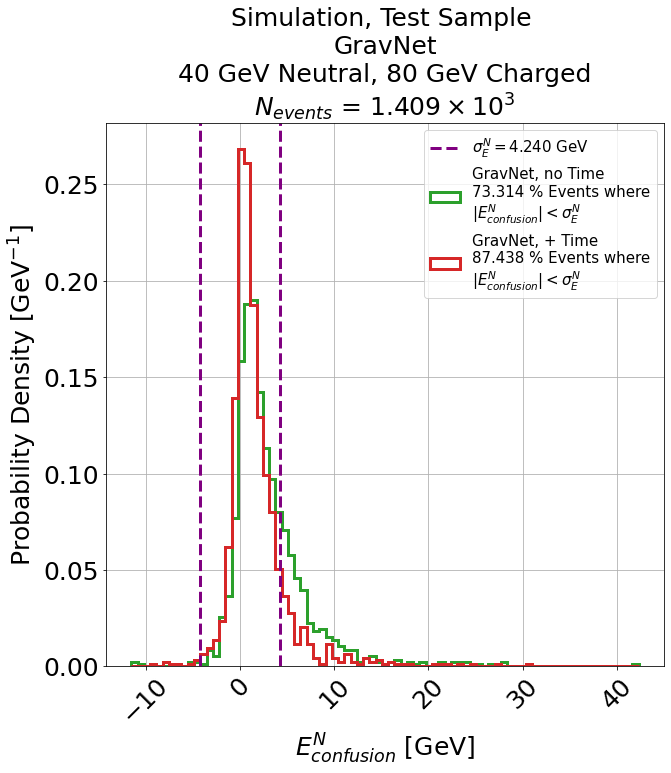}}

    \caption{ Figs.~\ref{fig:EConf_GN_rightskew_sim} and \ref{fig:EConf_GN_leftskew_sim} shows a \qty{40}{\giga \electronvolt} neutral shower separated from a \qty{10}{\giga \electronvolt} and \qty{80}{\giga \electronvolt} charged shower. The green and red lines indicate the models trained without and with timing information, respectively. The purple dashed lines indicate the resolution of the AHCAL calorimeter in simulation.}
    \label{fig:EConfusion_skew_examples}
\end{figure*}

Two example distributions from \texttt{GravNet} are presented in Fig.~\ref{fig:EConfusion_skew_examples} for further illustration.

\paragraph{2018 June Testbeam Data} 

\begin{figure*}
    \centering

    \subfloat[ \label{fig:Lin_GN_data_tos}]{\includegraphics[width=0.42\linewidth]{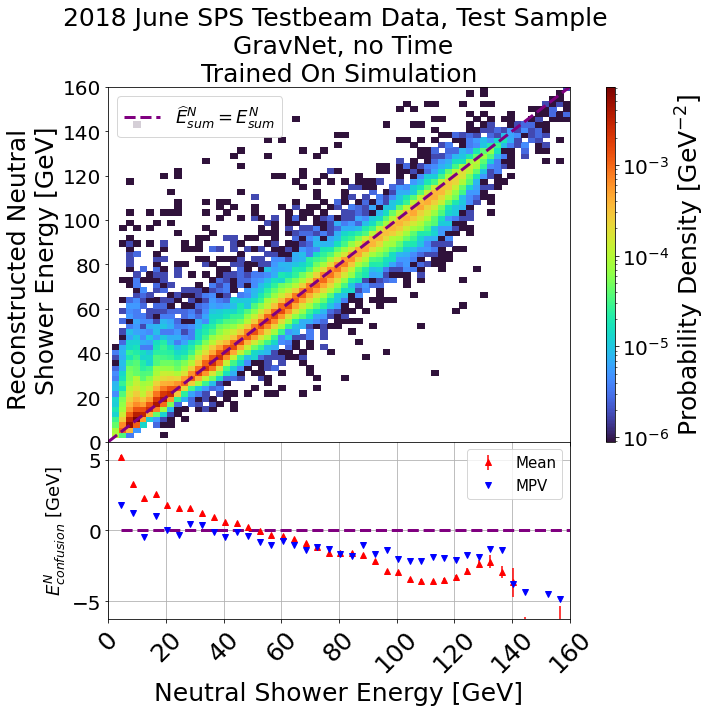}}
    \hfill
    \subfloat[\label{fig:Lin_GN_data_tod}]{\includegraphics[width=0.42\linewidth]{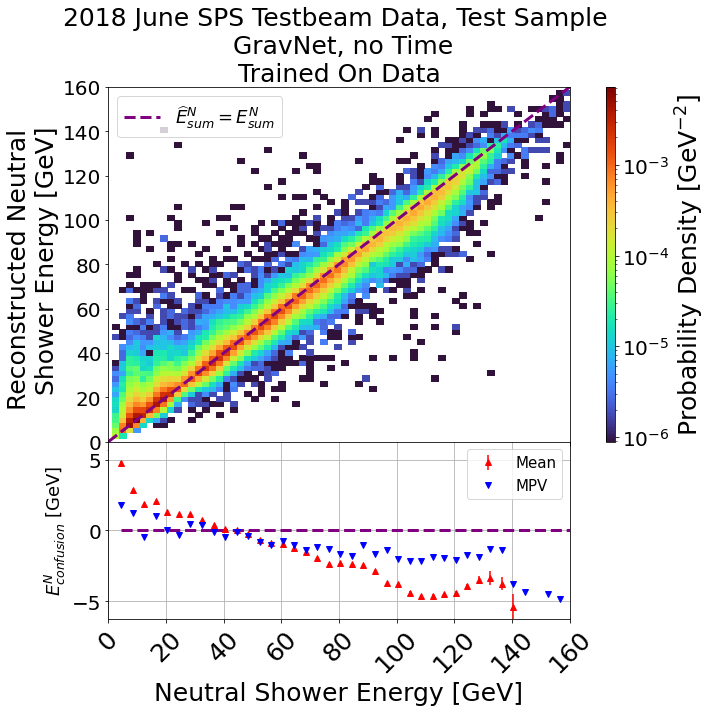}}

    \caption{Figs.~\ref{fig:Lin_GN_data_tos} and \ref{fig:Lin_GN_data_tod} show the joint distributions of the predicted and true reconstructed neutral shower response for \texttt{GravNet} trained on simulation and data and applied to the test sample of data, respectively. Else, as in Fig.~\ref{fig:Lin_sim}.}
    \label{fig:Lin_data}
\end{figure*}

\begin{figure*}
    \centering

    \subfloat[ \label{fig:Lin_GN_sim}]{\includegraphics[width=0.42\linewidth]{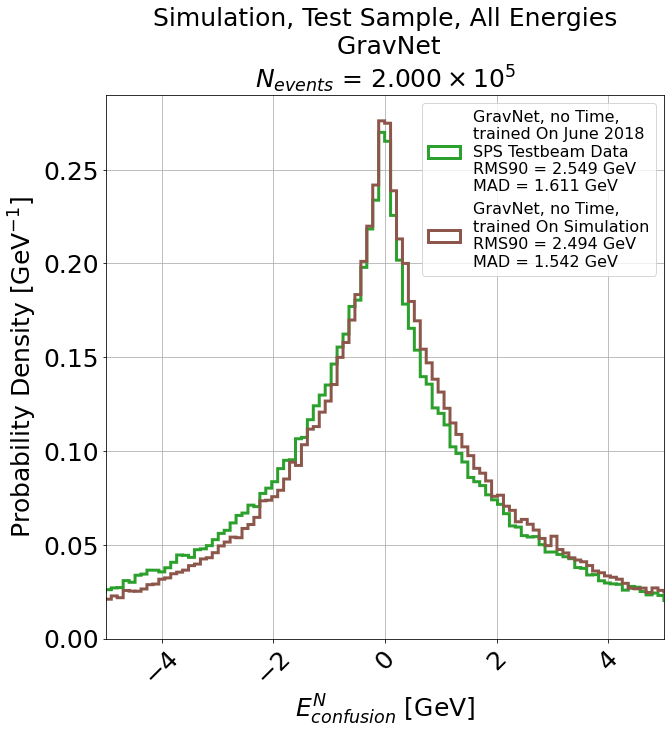}}
 
    \caption{The distributions of the neutral confusion energy for \texttt{GravNet} applied to the test sample of data. The green and brown lines indicate the models trained on simulation and data, respectively. Else, as in Fig.~\ref{fig:EConfusion_sim}.}
    \label{fig:EConfusion_data}
\end{figure*}

\begin{figure*}
    \centering

    \subfloat[ \label{fig:Skew_GN_data_tos}]{\includegraphics[width=0.4\linewidth]{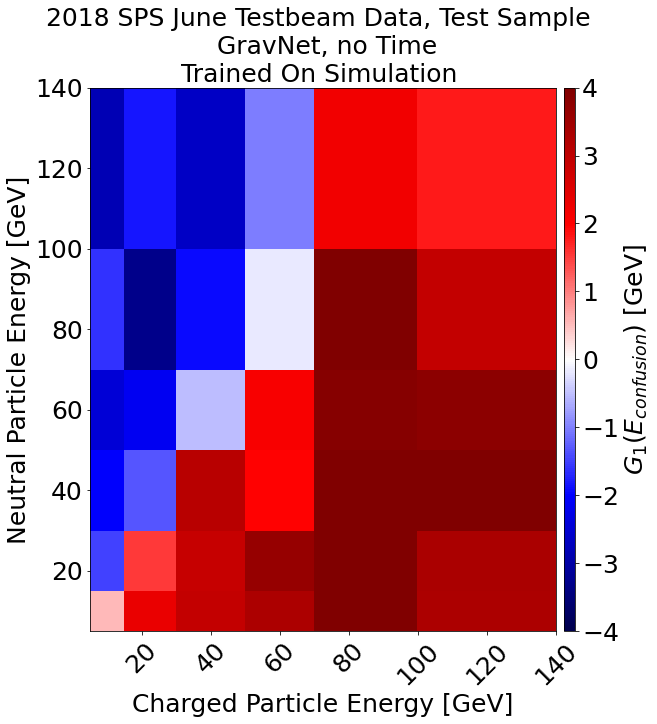}}
    \hfill
    \subfloat[\label{fig:Skew_GN_data_tod}]{\includegraphics[width=0.4\linewidth]{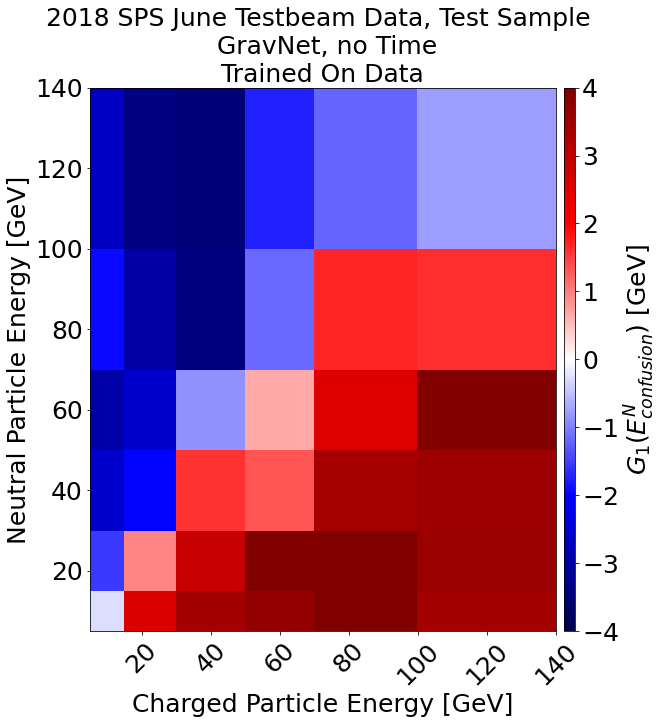}}

    \caption{Figs.~\ref{fig:Skew_GN_data_tos} and \ref{fig:Skew_GN_data_tod} show the skewness of the neutral confusion energy resolution gives the uncertainty on the hadron shower energy. Else, as in Fig.~\ref{fig:Skew_sim}.}
    \label{fig:Skew_data}
\end{figure*}

The reconstruction quality, overall confusion distributions and skewness shown in Figs.~\ref{fig:Lin_sim}, \ref{fig:EConfusion_sim} and \ref{fig:Skew_sim} are presented using the \texttt{GravNet} model evaluated on data in Figs.~\ref{fig:Lin_data}, \ref{fig:EConfusion_data} and \ref{fig:Skew_data}. Two models are assessed: the model trained on simulation and a separate model trained on data. 

The similarity of Figs.~\ref{fig:Lin_GN_data_tos} and \ref{fig:Lin_GN_data_tod} and the green and brown lines in Fig.~\ref{fig:EConfusion_data} indicate that the performance of the model trained on simulation and data achieve similar levels of confusion. This means models can be trained on simulation and applied to testbeam data with minimal difference in performance. Fig.~\ref{fig:Skew_data} shows the same behaviour of the skewness as previously discussed in Fig.~\ref{fig:Skew_sim}, indicating the behaviour is consistent in simulation and data. The apparent difference in the magnitude of the values between these figures results from the skewness being sensitive to outliers, and no appreciable difference in the shape of the confusion energy distributions of simulation and data is observed.

\subsection{Fraction of Energy Reconstructed Within Calorimeter Resolution}


For perfect separation of $Q$ and $N$, the energy resolution gives the uncertainty on the hadron shower energy. The performance of shower separation can, therefore, be quantified by the fraction of showers with reconstructed energy within one standard deviation from the true one, where the standard deviation is the calorimeter resolution at a given energy, defined $f_{\mathrm{rec}}$ and given by Eq.~\ref{eq:f_rec}: 

\begin{equation}
f_{\text {rec}} = \frac{N_{\left|E_{\text {confusion }}^N\right|<\sigma_E}}{N_{\text{events }}}
\label{eq:f_rec}
\end{equation}

where $N_{\text{events}}$ are the total number of showers in a studied subsample of the test dataset, $|E^{N}_{\text{confusion}}|<\sigma_{E}$ is the condition to be satisfied, $N_{|E^{N}_{\text{confusion}}|<\sigma_{E}}$ are the number of showers satisfying that condition and $\sigma_{E}$ is the calorimeter resolution, given by Eq.~\ref{eq:CaloSigma}:

\begin{equation}
   \sigma_{E} = a\cdot \sqrt{E} \oplus b \cdot E
   \label{eq:CaloSigma}
\end{equation}

where $E$ is particle energy, $a$ describes the combined sampling and stochastic fluctuations experienced by the calorimeter and $b$ the quality of detector calibration, non-uniformities in the signal collection, imperfections in calorimeter construction, etc., and $\oplus$ is an addition in quadrature.

The resolution terms $a$ and $b$ has been measured for simulation and data in Ref~\cite{SCNet} with $a_{\mathrm{sim}} = \qty[separate-uncertainty=true]{49.5 \pm 0.4}{\percent \per \sqrt{\giga \electronvolt}}$,  $b_{\mathrm{sim}} = \qty[separate-uncertainty=true]{7.1 \pm 0.1}{\percent}$ and $a_{\mathrm{data}}=\qty[separate-uncertainty=true]{56.1 \pm 0.7}{\percent \per \sqrt{\giga \electronvolt}}$, $b_{\mathrm{data}}=\qty[separate-uncertainty=true]{6.1 \pm 0.1}{\percent}$.

\begin{figure*}
    \centering

    \subfloat[ \label{fig:frec_PN_nt_sim}]{\includegraphics[width=0.33\textwidth]{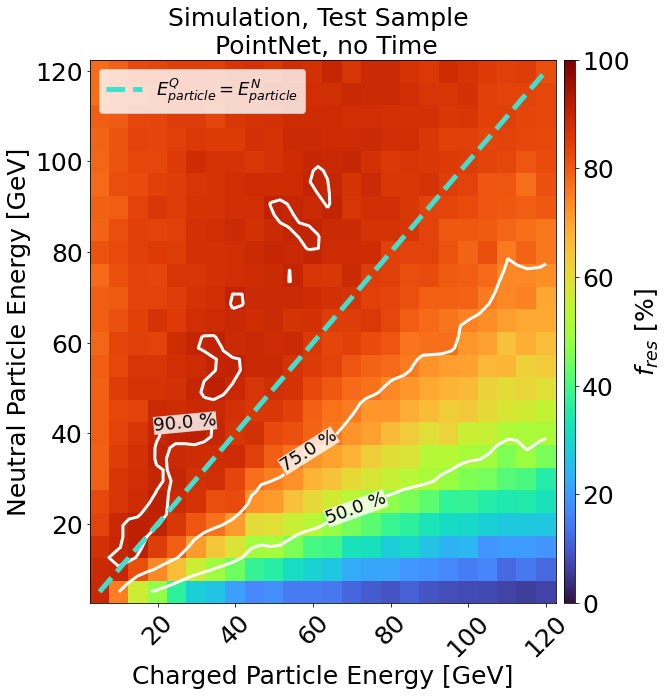}}
    \hfill
    \subfloat[\label{fig:frec_PN_wt_sim}]{\includegraphics[width=0.33\textwidth]{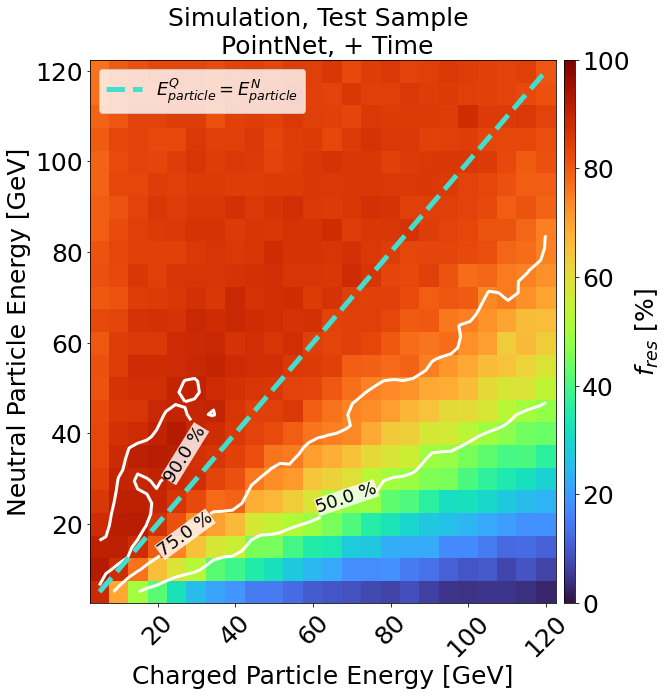}}
    \hfill
    \subfloat[ \label{fig:frec_PN_ratio}]{\includegraphics[width=0.33\textwidth]{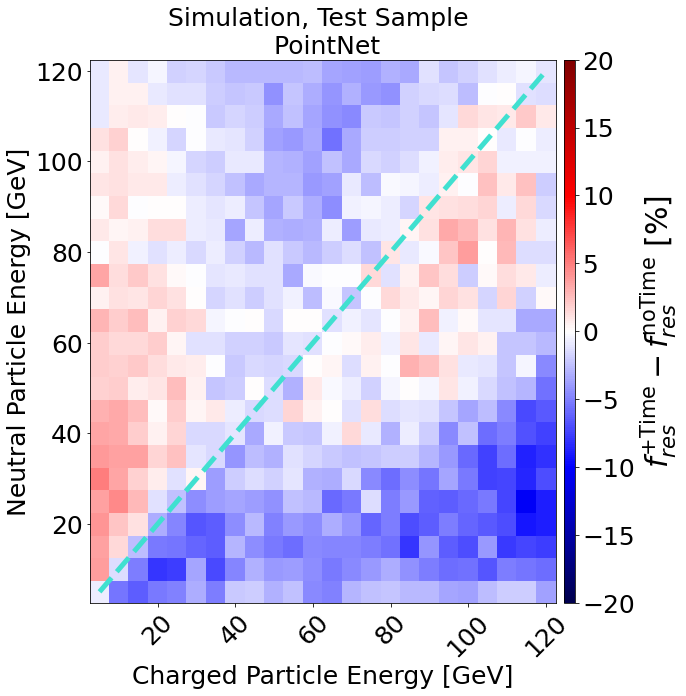}}
    
    \subfloat[ \label{fig:frec_DGCNN_nt_sim}]{\includegraphics[width=0.33\textwidth]{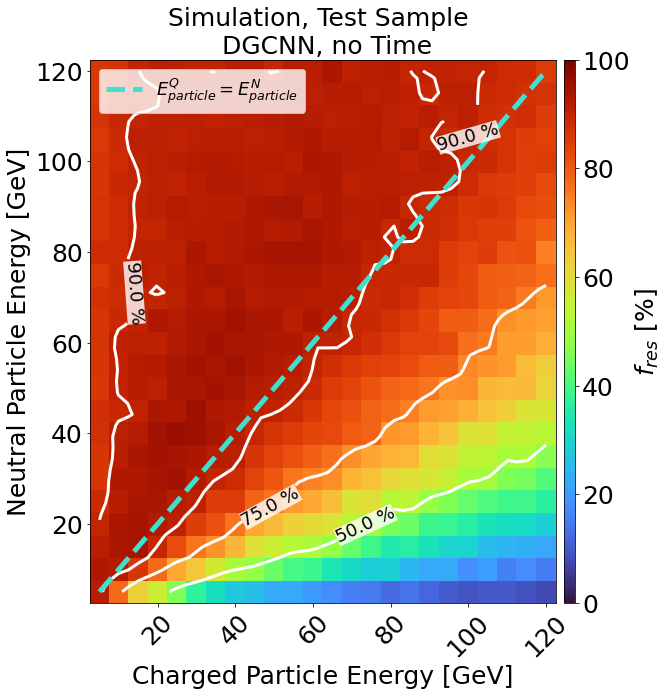}}
    \hfill
    \subfloat[\label{fig:frec_DGCNN_wt_sim}]{\includegraphics[width=0.33\textwidth]{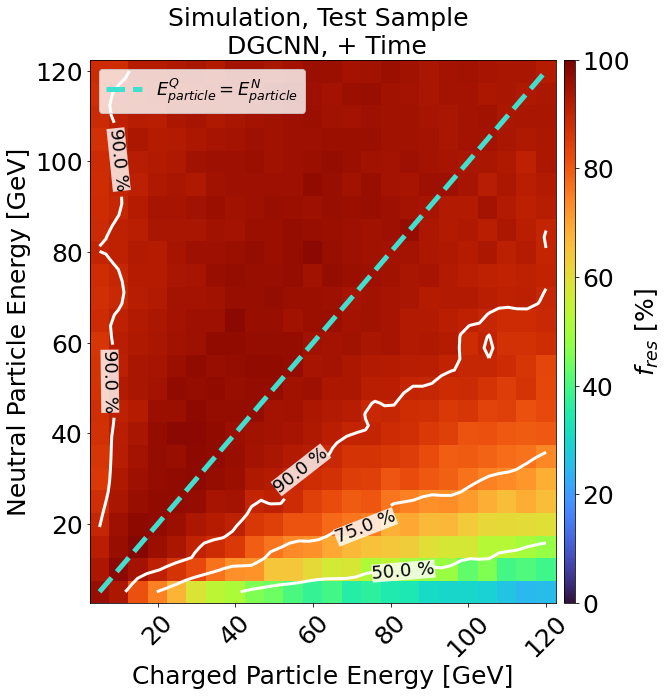}}
    \hfill
    \subfloat[ \label{fig:frec_DGCNN_ratio}]{\includegraphics[width=0.33\textwidth]{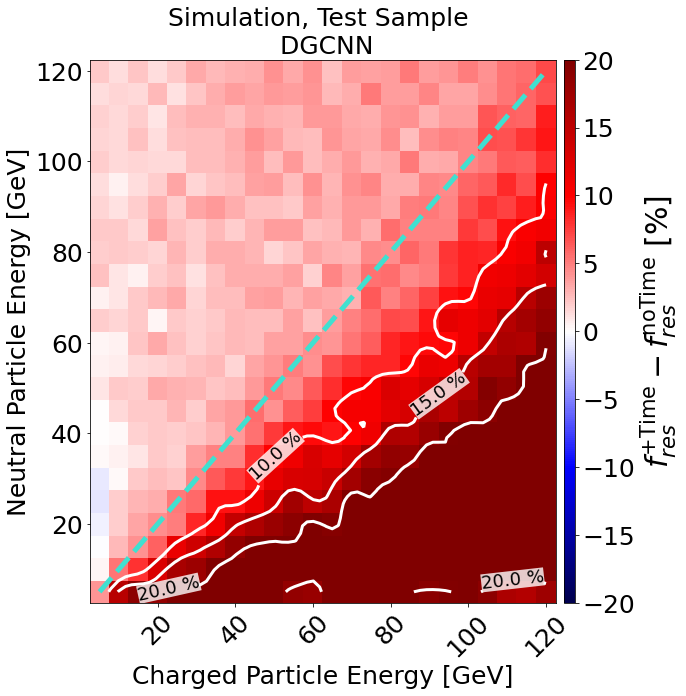}}

    \subfloat[ \label{fig:frec_GN_nt_sim}]{\includegraphics[width=0.33\textwidth]{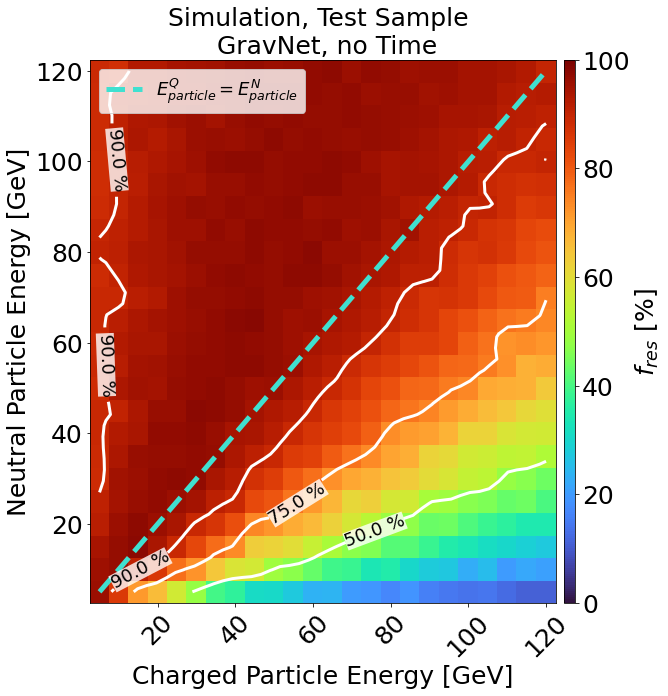}}
    \hfill
    \subfloat[\label{fig:frec_GN_wt_sim}]{\includegraphics[width=0.33\textwidth]{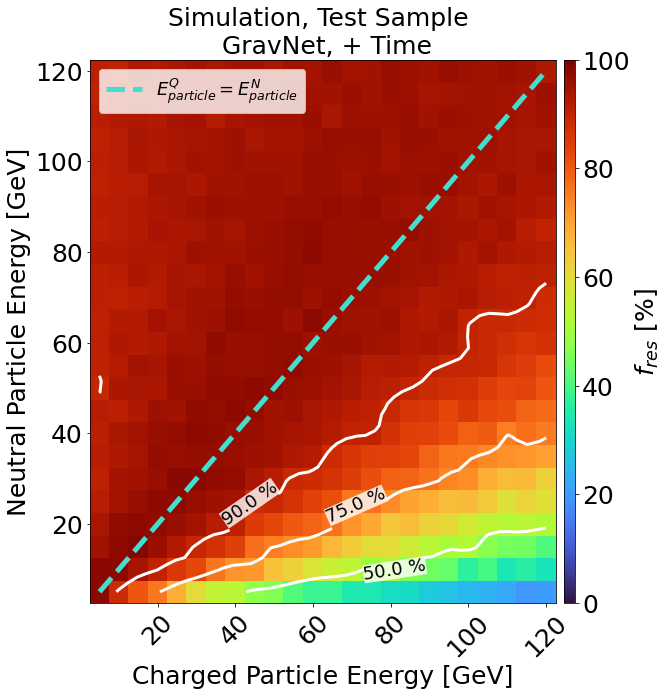}}
    \hfill
    \subfloat[ \label{fig:frec_GN_ratio}]{\includegraphics[width=0.33\textwidth]{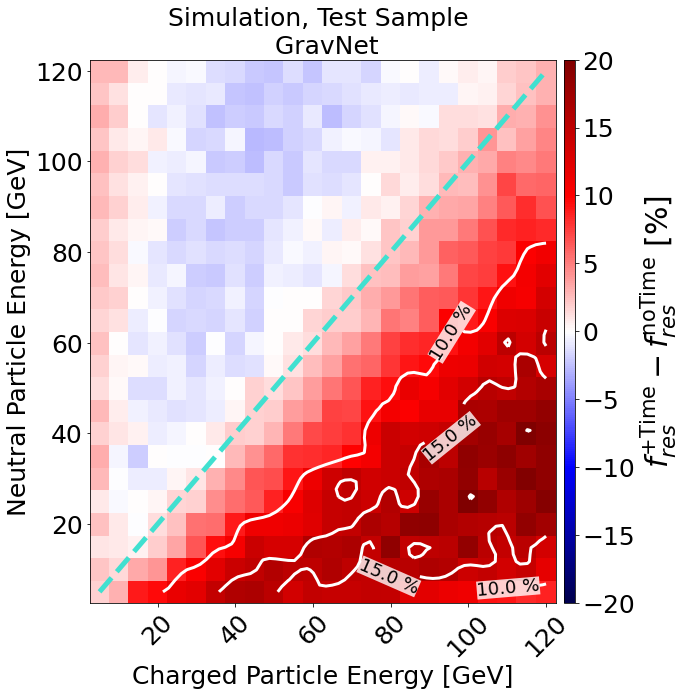}}

    \caption{Figures \ref{fig:frec_PN_nt_sim}, \ref{fig:frec_DGCNN_nt_sim}, \ref{fig:frec_GN_nt_sim} and Figures \ref{fig:frec_PN_wt_sim}, \ref{fig:frec_DGCNN_wt_sim},
    \ref{fig:frec_GN_wt_sim} show the matrices of the fraction of showers with energy reconstructed within the calorimeter resolution as a function of the charged and neutral particle energy, for \texttt{PointNet}, \texttt{DGCNN} and \texttt{GravNet}, respectively, where red to blue indicates a higher to lower percentage of events. The turquoise dashed line indicates $E^{Q}_{\text{particle}} =E^{N}_{\text{particle}}$, while white lines indicate contours. Figures \ref{fig:frec_PN_ratio}, \ref{fig:frec_DGCNN_ratio} and \ref{fig:frec_GN_ratio} indicate the ratios of the fractions for models trained with and without timing information.}
    \label{fig:f_rec_sim}
\end{figure*}

\begin{figure*}
    \centering



    \subfloat[ \label{fig:GN_Sim_10Q10N}]{\includegraphics[width=0.49\linewidth]{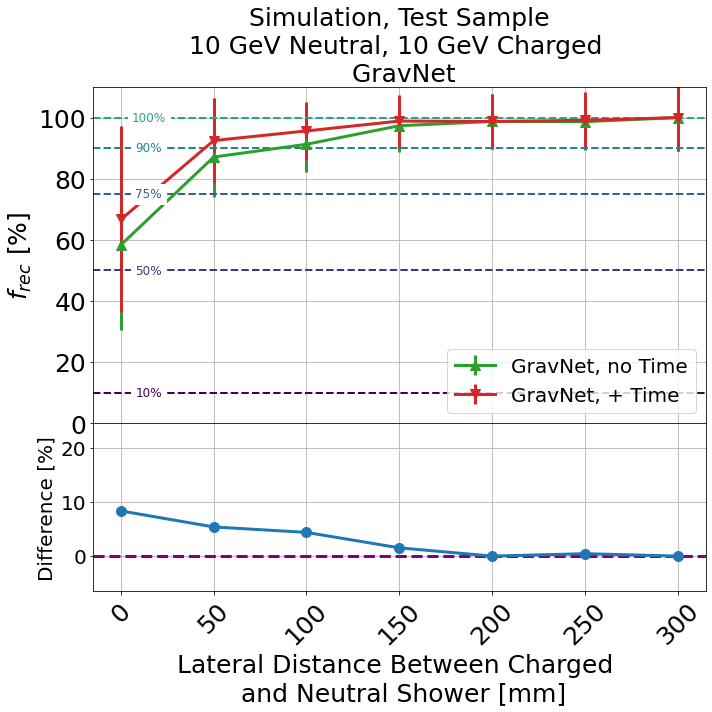}}
    \hfill
    \subfloat[\label{fig:GN_Sim_30Q10N}]{\includegraphics[width=0.49\linewidth]{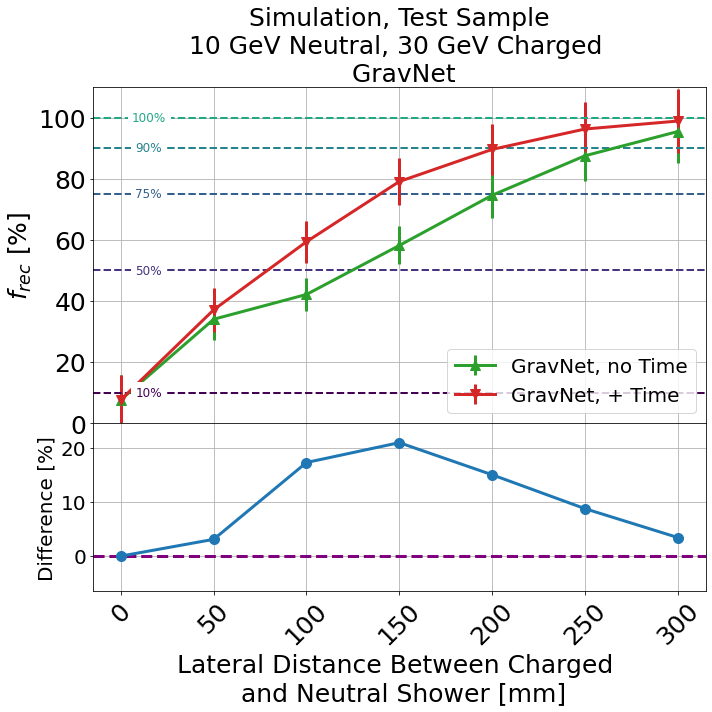}}
     \caption{Figs.~\ref{fig:GN_Sim_10Q10N} and \ref{fig:GN_Sim_30Q10N} show $f_{\mathrm{rec}}$ (see Eq.~\ref{eq:f_rec}) as a function of the distance between showers $Q$ and $N$ in \unit{\milli \meter} for the test sample of simulation with the \texttt{GravNet} model, with $E_{Q}=E_{N}=\qty{10}{\giga \electronvolt}$ and $E_{Q}=\qty{30}{\giga \electronvolt}$, $E_{N}=\qty{10}{\giga \electronvolt}$, respectively. The red and green lines indicate the models trained with and without time, respectively. The subplots with the blue lines indicate the additional fraction of showers, in percent, reconstructed by the model trained with time than without.}
    \label{fig:RSep_frec_sim}

\end{figure*}

\paragraph{Simulation} $f_{\mathrm{rec}}$ was calculated for the test sample of simulation for all models under test. The results are shown as a function of the charged and neutral particle energy in Fig.~\ref{fig:f_rec_sim}.

Figures \ref{fig:frec_PN_nt_sim}, \ref{fig:frec_DGCNN_nt_sim}, and \ref{fig:frec_GN_nt_sim}, along with Figures \ref{fig:frec_PN_wt_sim}, \ref{fig:frec_DGCNN_wt_sim}, and \ref{fig:frec_GN_wt_sim}, reveal an asymmetry in shower reconstruction performance based on whether the charged particle energy surpasses that of the neutral particle. When the neutral hadron possesses more energy than the charged one, from \qty{80}{\percent} to well over \qty{90}{\percent} of showers are reconstructed within the calorimeter resolution for \texttt{DGCNN} and \texttt{GravNet}. However, performance deteriorates in the opposite scenario. This result can be attributed to the relative magnitude of $\sigma_{E}$ compared to $E_{\text{confusion}}$. To clarify, confusion was observed to be about the same for a particular set of particle energies, regardless of whether it was $Q$ or $N$ that deposited more or less energy. However, the $\sigma_{E}$ of the calorimeter is smaller when $N$ has less energy, which means $f_{\text{rec}}$ increases in that case. The opposite is true when $N$ has more energy.


Figs.~\ref{fig:frec_DGCNN_ratio} and \ref{fig:frec_GN_ratio} demonstrate that \texttt{DGCNN} and \texttt{GravNet} achieve improvements of up to an additional 15-\qty{20}{\percent} of showers with the inclusion of timing information where the energy of the charged shower surpasses the neutral. The red region below the cyan equality line indicates this. Conversely, no notable improvement is observed in the opposite scenario. \texttt{PointNet}, as depicted in Fig.~\ref{fig:frec_PN_ratio}, does not exhibit such enhancements. This result indicates that a more sophisticated clustering method is obtained using timing information than without, particularly in the absence of track information.

Fig.~\ref{fig:RSep_frec_sim} shows the correlation between $f_{\mathrm{rec}}$ and the lateral distance between $Q$ and $N$, highlighting the enhancement achieved by incorporating timing data into the \texttt{GravNet} network for two benchmark PF separation scenarios. Specifically, Fig.~\ref{fig:GN_Sim_10Q10N} shows cases where the particle energies are $E_{Q}=E_{N}=\qty{10}{\giga \electronvolt}$, while Figure \ref{fig:GN_Sim_30Q10N} represents $E_{Q}=\qty{30}{\giga \electronvolt}$ and $E_{N}=\qty{10}{\giga \electronvolt}$. Fig.~\ref{fig:GN_Sim_10Q10N} shows marginal and statistically insignificant improvement by including timing information when the energies of $Q$ and $N$ are comparable. In contrast, Figure \ref{fig:GN_Sim_30Q10N} demonstrates a notable and statistically significant improvement, particularly at distances ranging from 50- \qty{250}{\milli \meter}, with an increase of up to \qty{20}{\percent} in the number of showers reconstructed within the resolution. Once again, these results highlight the relevance of the temporal aspect of the AHCAL to PF shower separation.

\begin{figure*}
    \centering

    \subfloat[ \label{fig:frec_GN_nt_tos_data}]{\includegraphics[width=0.33\textwidth]{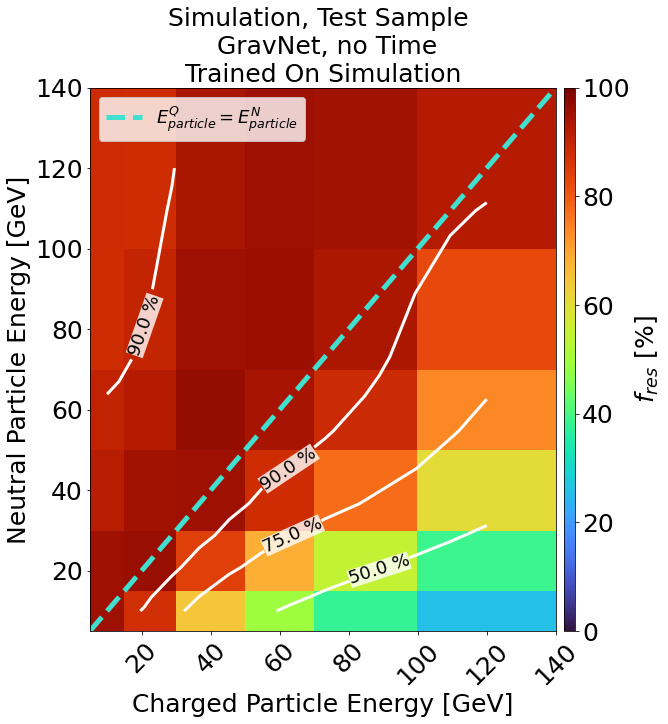}}
    \hfill
    \subfloat[\label{fig:frec_GN_nt_tod_data}]{\includegraphics[width=0.33\textwidth]{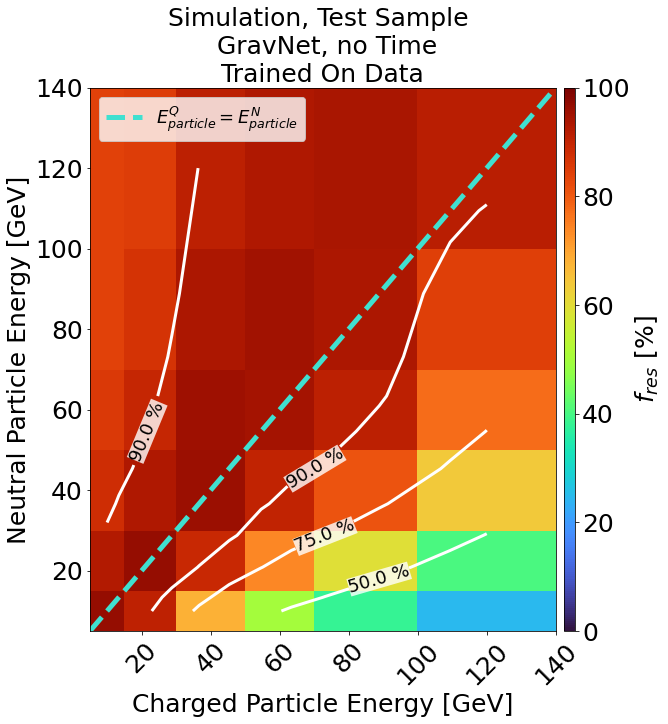}}
    \hfill
    \subfloat[ \label{fig:frec_GN_ratio_data}]{\includegraphics[width=0.33\textwidth]{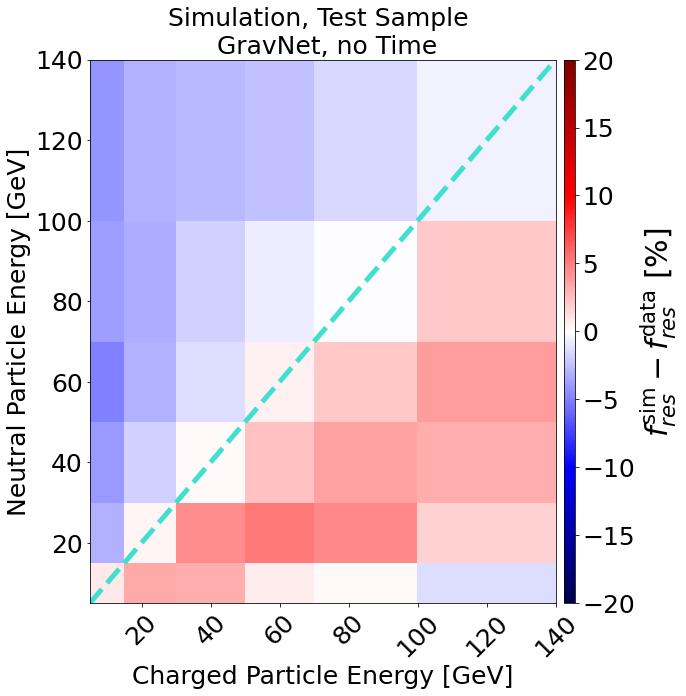}}

    \caption{Figs.~\ref{fig:frec_GN_nt_tos_data} and \ref{fig:frec_GN_nt_tod_data} show the matrices of the fraction of showers with energy reconstructed within the calorimeter resolution as a function of the charged and neutral particle energy for \texttt{GravNet}, trained on simulation and data, respectively. Fig.~\ref{fig:frec_GN_ratio_data}
indicates the ratios of the fractions. Else, as in Fig.~\ref{fig:f_rec_sim}.}
    \label{fig:f_rec_data}
\end{figure*}

\begin{figure*}
    \centering

    \subfloat[ \label{fig:GN_Data_10Q10N}]{\includegraphics[width=0.49\linewidth]{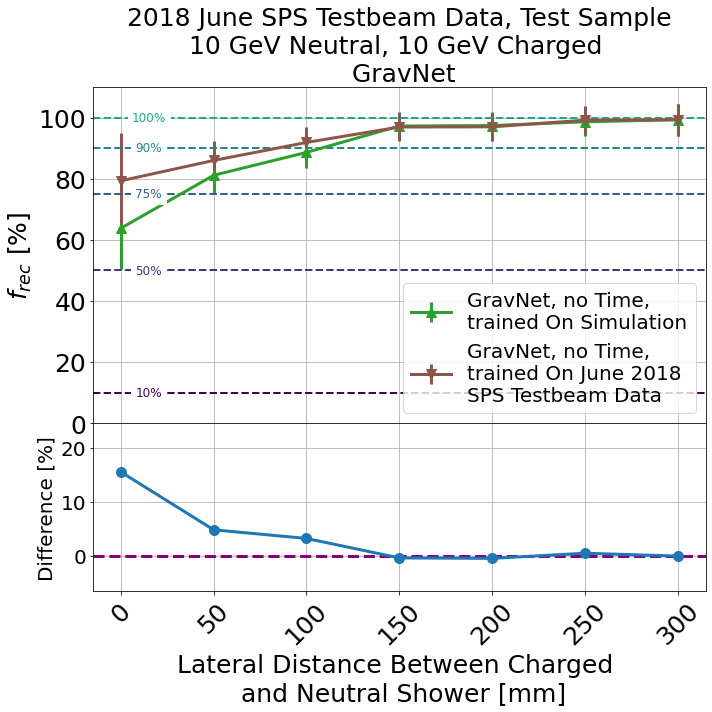}}
    \hfill
    \subfloat[\label{fig:GN_Data_30Q10N}]{\includegraphics[width=0.49\linewidth]{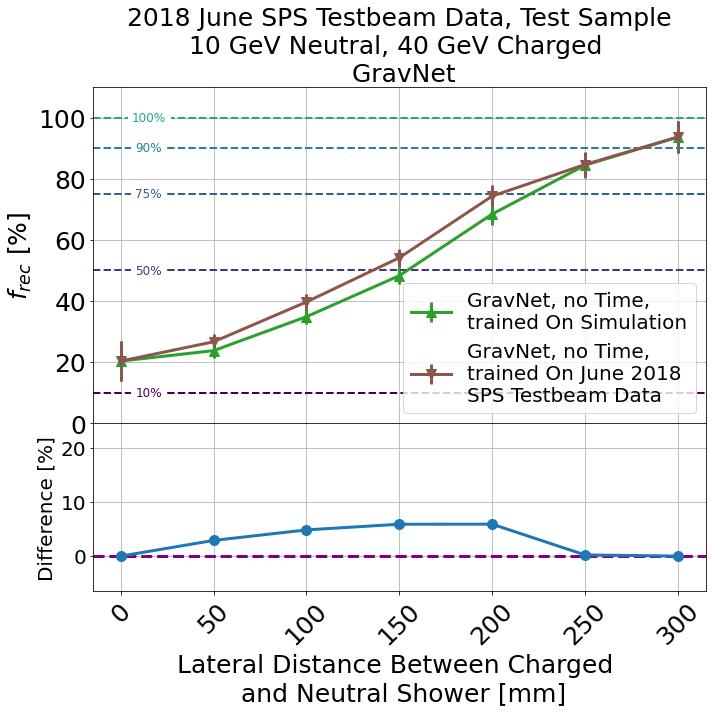}}
  
     \caption{Figs.~\ref{fig:GN_Data_10Q10N} and \ref{fig:GN_Data_30Q10N} show $f_{\mathrm{rec}}$ (see Eq.~\ref{eq:f_rec}) as a function of the distance between showers $Q$ and $N$ in \unit{\milli \meter} for the test sample of data with the \texttt{GravNet} model, with $E_{Q}=E_{N}=\qty{10}{\giga \electronvolt}$ and $E_{Q}=\qty{40}{\giga \electronvolt}$, $E_{N}=\qty{10}{\giga \electronvolt}$, respectively. The green and brown lines indicate the models trained on simulation and data, respectively. Else, as in Fig.~\ref{fig:RSep_frec_sim}.}
    \label{fig:RSep_frec_data}

\end{figure*}

\paragraph{2018 June Testbeam Data}
As in Fig.~\ref{fig:f_rec_sim}, the \texttt{GravNet} model trained on the training sample of simulation and data is evaluated on the test sample of data. The results are shown in Fig.~\ref{fig:f_rec_data}.

Figs.~\ref{fig:frec_GN_nt_tos_data} and \ref{fig:frec_GN_nt_tod_data} show the same asymmetry as presented in Fig.~\ref{fig:f_rec_sim}. This result indicates the consistent effect between training on simulation or data showers.

Fig.~\ref{fig:frec_GN_ratio_data} indicates that the fraction of showers reconstructed within the resolution of the AHCAL by the \texttt{GravNet} network trained on simulation and data varies by no more than around \qty{5}{\percent}. This result means that the performance of the neural networks is not strongly related to the choice to use simulation or data for training the models.  

Fig.~\ref{fig:RSep_frec_data} the same information as Fig.~\ref{fig:RSep_frec_sim} but for data and with $E_{Q}=\qty{40}{\giga \electronvolt}$, $E_{N}=\qty{10}{\giga \electronvolt}$ in Fig.~\ref{fig:GN_Data_30Q10N}, as a \qty{30}{\giga \electronvolt} sample was not used in the study. The similarity of of the green and brown lines indicate agreement with the previous statement. The difference in the trend between Figs.~\ref{fig:GN_Sim_30Q10N} and \ref{fig:GN_Data_30Q10N} can be attributed to the difference in energy between $Q$ and $N$ being larger in the latter case, resulting in more overall confusion.

\section{Conclusion}

Three published neural networks (\texttt{PointNet}, \texttt{DGCNN} and \texttt{GravNet}) were trained to separate a charged and neutral hadron shower with the AHCAL technological prototype to evaluate the shower separation performance of the calorimeter. The neural networks were trained to separate synthetic showers with two showers produced using a method to overlay two hadron showers from single showers. The position of a single shower was uniformly distributed with a fixed most-probable distance between showers and had a uniform energy distribution between 5 and \qty{120}{\giga \electronvolt}. Simulation and data were used, as well as timing information in simulation with \qty{100}{\pico \second} resolution. The networks were evaluated, and the results were studied.

Firstly, in simulation, it was observed that \texttt{PointNet} did not improve resolution using timing information. By contrast, \texttt{DGCNN} and \texttt{GravNet} observed a significant reduction in confusion using timing information. For the best-performing neural network (\texttt{GravNet}) this corresponded to a reduction of the MAD by around \qty{23}{\percent}. This result was speculatively attributed to the improved sensitivity of \texttt{GravNet} and \texttt{DGCNN} to 'local energy density' compared to \texttt{PointNet}, which does not exploit this information by design. 

Secondly, all models exhibited asymmetry in the confusion energy distribution, with a tendency to allocate more energy to the less energetic of the two showers rather than the more energetic one. This result was attributed to an 'altruistic' clustering method that produces similar distributions to those observed in similar studies using Pandora PFA. 

Thirdly, all models were found to reconstruct 80-\qty{90}{\ per cent} of showers within the calorimeter resolution where the neutral particle had more energy than the charged one and rapidly degraded in the opposite case. This behaviour was attributed to the better performance achievable when there is a significant disparity between track position and the centre-of-gravity of the most energetic shower, which is rarely the case when the charged shower has more energy than the neutral. Timing information was found to explicitly increase the number of showers reconstructed correctly of the latter case by 15-\qty{20}{\percent} using \texttt{DGCNN} and \texttt{GravNet}, and motivates the temporal component of the AHCAL calorimeter exploited by graph neural networks. 

Finally, the studies made on simulation were repeated for the 2018 June SPS Testbeam data, and the simulation-trained and data-trained models were compared. Regarding performance and properties, almost no difference was observed between the model trained on simulation and the model trained on data. Note that no timing information was available in the study of the test beam data. Therefore, timing was only applied in the simulation studies.

This study suggests that the AHCAL is a highly effective PF calorimeter whose performance can be enhanced using timing information with a resolution of \qty{100}{\pico \second}. It also concludes that shower separation algorithms can be trained on simulation and applied directly to experimental data with similar performance.

\clearpage 
\appendix

\section{Appendix}

\subsection{Summaries of Neural Networks} \label{sec:SummaryNN}

\paragraph{\texttt{PointNet}}

 The paper's implementation is based on Ref~\cite{pointnet_implementation}. Firstly, the hit indices of the shower ($I_{\mathrm{hit}}$, $J_{\mathrm{hit}}$, $K_{\mathrm{hit}}$) pass through a 'transformation network' (T-Net, see Ref~\cite{pointnet}), which produces an affine transformation matrix. The T-Net includes an upsampling module with four sequential 1D fully connected layers (64, 128, 256, and 512 channels), using mean, variance, and maximum pooling and a downsampling module with three sequential 1D fully connected layers (256, 128, and 9 channels), with batch normalization and leaky ReLU activation. The output matrix is multiplied by the input. Additional features of the hadron shower relevant to clustering ($\log{E_{\mathrm{hit}}}$, $\operatorname{arcsinh}t_{\mathrm{hit}}$, $I_{\mathrm{track}}$, $J_{\mathrm{track}}$) are concatenated and upscaled by a 1D fully-connected layer with 64 channels. Then, the output passes through a second T-Net of the same structure as the 3D Transform with $64 \times 64$ layers. The output, denoted as $A$, is initialised as the identity matrix and is used with mean, variance and maximum pooling. The remaining hit energy and track energy information are concatenated to the output. The final module includes four sequential 1D fully connected layers (512, 256, 128, and 64 channels), with batch normalisation, leaky ReLU activation, and dropout in the last layer. The final layer outputs reconstructed energy fractions for each hit in the shower, using Softmax activation. The matrix $A$ is regularized during training to remain symmetric to ensure affine transformation. A diagram indicating the model's design is shown in Appendix Fig.~\ref{fig:ShowerSep_PointNet}.

\paragraph{\texttt{DGCNN}} The paper's implementation, based on Ref~\cite{dgcnn_implementation}, utilizes four \texttt{EdgeConv} operators (see Ref~\cite{dgcnn}). The input is concatenated with its mean in each operation and passed through a $k$-NN clustering, followed by two fully connected 2D layers with 64 channels. The mean, variance, and maximum are calculated over the clusters and passed through a 1D fully-connected layer with 64 channels, along with batch normalisation and leaky ReLU activation for all layers. The first \texttt{EdgeConv} operator uses only hit indices of the shower ($I_{\mathrm{hit}}$, $J_{\mathrm{hit}}$, $K_{\mathrm{hit}}$). Additional 'features' of the hadron shower relevant to clustering ($\log{E_{\mathrm{hit}}}$, $\operatorname{arcsinh}t_{\mathrm{hit}}$, $I_{\mathrm{track}}$, $J_{\mathrm{track}}$) are later concatenated to the output of the first \texttt{EdgeConv} operator before applying the second operator. At each stage, the output from each \texttt{EdgeConv} operator is recorded and later concatenated into a single tensor. Then, one shared 1D fully connected layer with 1024 channels, batch normalization, and leaky ReLU activation condenses the features learned at the clustering stage. Mean, variance and maximum pooling over the points activate the points, with remaining hit energy and track energy information ($E_{\mathrm{hit}}$, $E^{Q}_{\mathrm{tr}}$) concatenated to the output. The final stage of the network includes four sequential 1D fully connected layers (512, 256, 128, and 64 channels), with batch normalization, leaky ReLU activation, and dropout in the last layer. The final layer produces two outputs, one for $\widehat{f}_{\mathrm{hit}}^{Q}$ and one for $\widehat{f}_{\mathrm{hit}}^{N}$, using Softmax activation, resulting in reconstructed energy fractions for each hit in the shower. A diagram indicating the model's design is shown in Appendix Fig.~\ref{fig:ShowerSep_EdgeConvDGCNN}.

\paragraph{\texttt{GravNet}}

The paper's implementation, based on Refs.~\cite{dgcnn_implementation} and \cite{gravnet_implementation}, replaces \texttt{EdgeConv} layers with \texttt{GravNet} layers in the model. Before each \texttt{GravNet} layer, the mean is concatenated to the input and passed through two fully connected 1D layers with batch normalization and leaky ReLU activation, each with 64 channels.

In this paper, the \texttt{GravNet} layer operation starts by projecting points to a low-dimensional clustering space, $s$, and then to a high-dimensional learned feature space, $F_{LR}$. This process involves two individual 1D fully connected layers with 4 (5 if time is included) and 22 channels, respectively. Then, a $k$-NN cluster is found in the $s$-space, similar to \texttt{DGCNN}. Euclidean distances of the neighbours to the graph's origin are calculated for each cluster. These distances are scaled by a Gaussian potential with a hyperparameter 'potential strength' parameter $\gamma$, resulting in aggregated values across the cluster using maximum, mean, and variance. Finally, a 1D fully connected layer with 48 channels, batch normalization, and leaky ReLU activation condenses the aggregates to a single value. A diagram indicating the design of the model is shown in Appendix Fig.~\ref{fig:ShowerSep_GravNetConv}.

\clearpage

\begin{figure*}[!htb]
    \centering
    \includegraphics[width=\linewidth]{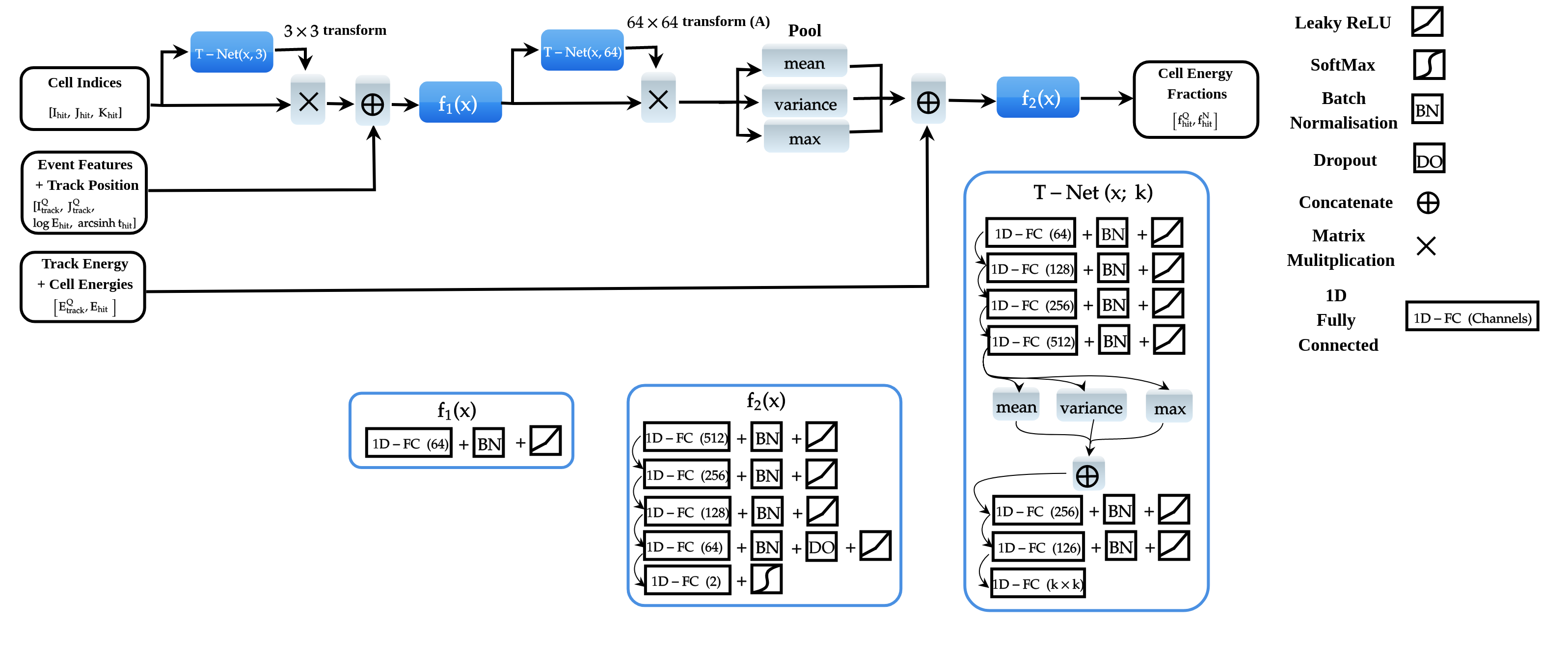}
    \caption[Illustration of \texttt{PointNet} Model Used For Shower Separation]{Diagram illustrating the \texttt{PointNet} implementation in this study. The black, blue and grey boxes indicate inputs and outputs, convolutional operations and general operations, respectively. Additional operations are specified on the right of the figure. Additionally, matrix multiplication is indicated as a $\times$ symbol, batch normalisation is denoted 'BN' and the Softmax activation is indicated in the legend.}
    \label{fig:ShowerSep_PointNet}
\end{figure*}

\begin{figure*}[!htb]
    \centering

     \subfloat[ \label{fig:ShowerSep_EdgeConv}]{\includegraphics[width=\linewidth]{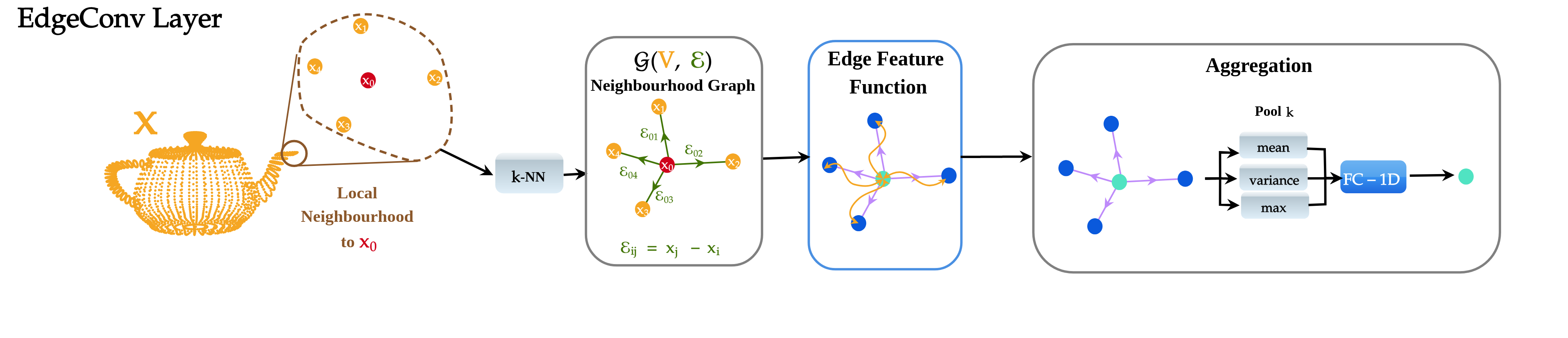}}

     \subfloat[ \label{fig:ShowerSep_DGCNN}]{\includegraphics[width=\linewidth]{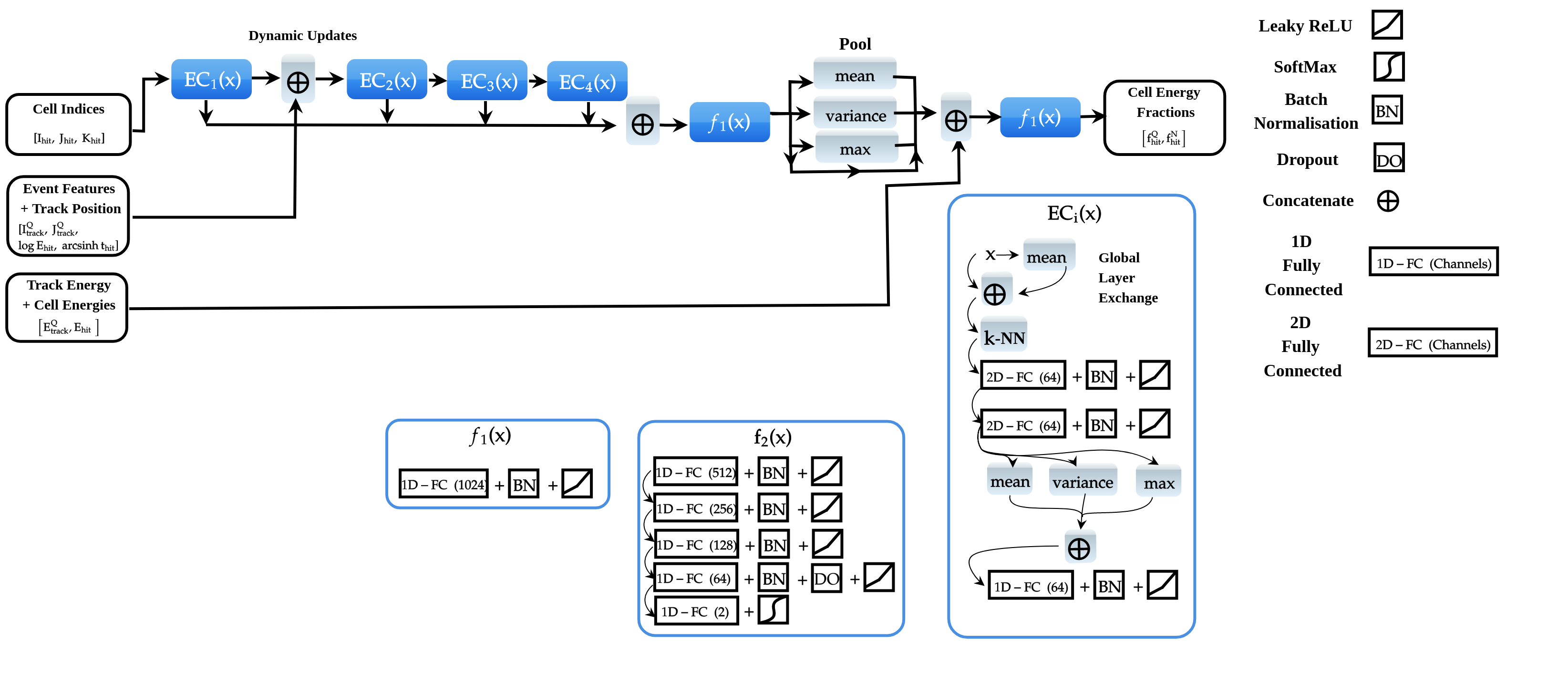}}

    \caption[Illustration of \texttt{DGCNN} Model and \texttt{EdgeConv} Layer Used For Shower Separation]{Figure \ref{fig:ShowerSep_EdgeConv} illustrates the \texttt{EdgeConv} operator. The orange Utah teapot indicates some 'point cloud', or distribution of points, $x$, with some underlying distribution, to be operated upon. The orange dots indicate vertices, $V$, of a local neighbourhood $k$-NN graph around the central red dot. The green arrows indicate the vectors (edges) between the central red dot and its neighbours, $\mathcal{E}$. The orange arrows indicate 'message-passing' between the vertices and the edges, which modifies the graph as indicated by the colour inversion. Figure \ref{fig:ShowerSep_DGCNN} shows a diagram illustrating the \texttt{DGCNN} implementation in this study. Else, as in Fig.~ \ref{fig:ShowerSep_PointNet}.}
    \label{fig:ShowerSep_EdgeConvDGCNN}
\end{figure*}

\begin{figure*}[!htb]
    \centering

     \subfloat[ \label{fig:ShowerSep_GravConv}]{\includegraphics[width=\linewidth]{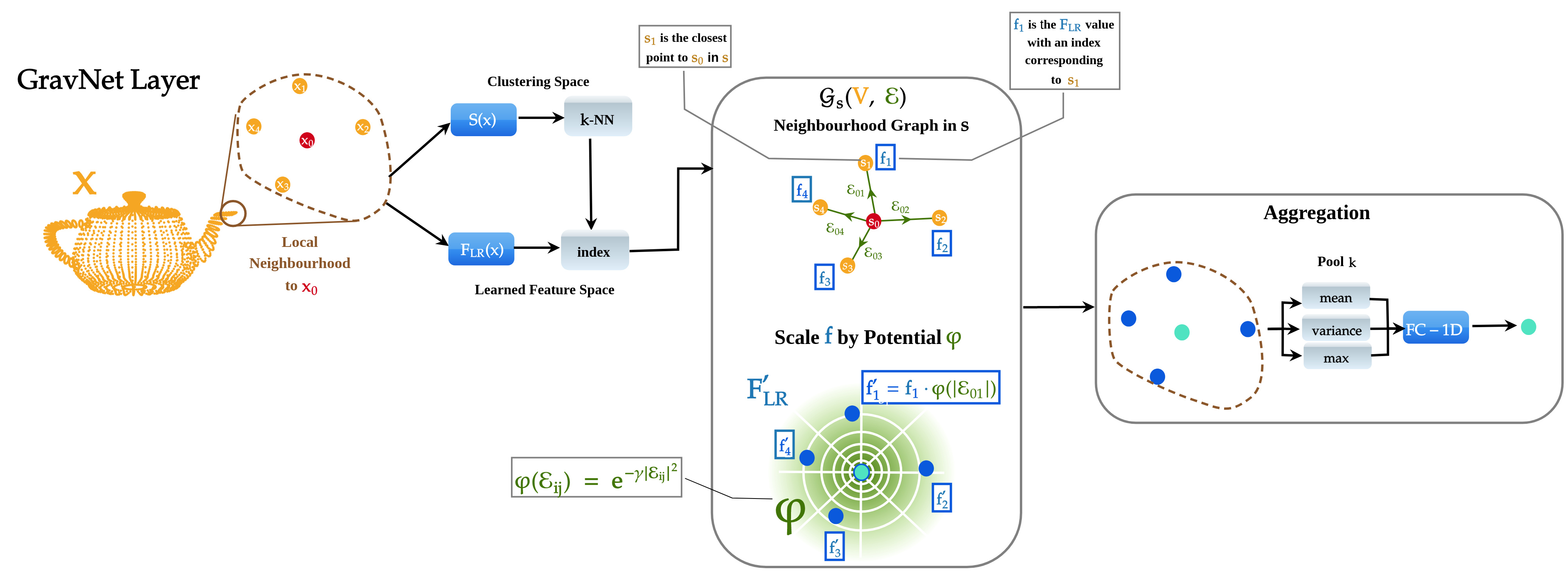}}

     \subfloat[ \label{fig:ShowerSep_GravNet}]{\includegraphics[width=\linewidth]{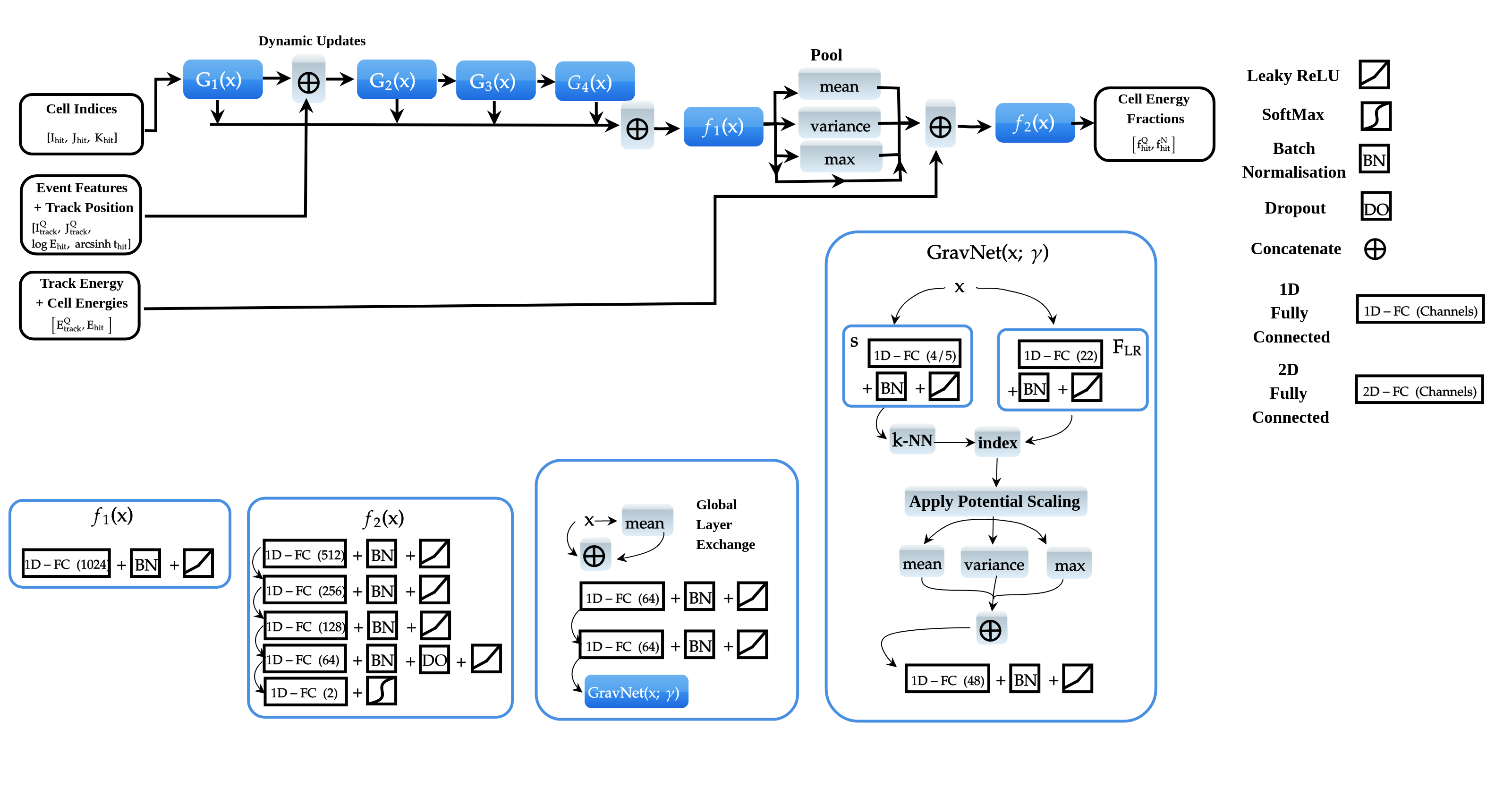}}

    \caption[Illustration of \texttt{DGCNN} Model and \texttt{EdgeConv} Layer Used For Shower Separation]{Fig.~\ref{fig:ShowerSep_GravConv} illustrates the \texttt{GravNet} operator. The orange Utah teapot indicates some 'point cloud', or distribution of points, $x$, with some underlying distribution, to be operated upon. The $S$ and $F_{LR}$ operations indicate the clustering and feature space representations learned by the neural network. The green arrows and $\mathcal{E}$ correspond to vectors between the red point and its orange neighbours, its norm indicating a distance in a potential well given by $\varphi$, indicated by the shaded green region. $f$ and $f^{'}$ indicate unscaled and potential-scaled features. Else as in Figure \ref{fig:ShowerSep_EdgeConv}.}
    \label{fig:ShowerSep_GravNetConv}
\end{figure*}

\FloatBarrier

\subsection{Validation of Synthetic Neutral Hadron Showers}\label{sec:ROCTest}

\begin{table*}
    \caption{Table \ref{tab:EventTable_BDT} shows a subsample of Table \ref{tab:EventTable} of \qty{40}{\giga \electronvolt} particles used for training the gradient-boosted decision tree to assess the performance of the synthetic neutral hadron showers produced via the MIP-track cut. Table \ref{tab:BDTHyperparams} shows the hyperparameters used for the CALICE PID gradient-boosted decision tree,  used in this paper to analyse the effectiveness of the MIP-track cut. Further information about these parameters can be found in \cite{lightGBM}.}
    \centering
    \begin{minipage}[t]{0.45\textwidth}
        \centering
        \subfloat[\label{tab:EventTable_BDT}]{
            \scriptsize
            \begin{tabular}{lrrr}
                \toprule
                & \multicolumn{3}{c}{Simulation} \\
                Hadron & Testing & Training & Validation \\
                \midrule
                $K^0_L$ & 11732 & 54800 & 11614 \\
                $\pi^-$ & 11411 & 53201 & 11530 \\
                \bottomrule
            \end{tabular}
        }
    \end{minipage}%
    \hfill
    \begin{minipage}[t]{0.45\textwidth}
        \centering
        \subfloat[\label{tab:BDTHyperparams}]{
            \scriptsize
            \begin{tabular}{ll}
                \toprule
                Hyperparameter & Value \\
                \midrule
                Objective & SoftMax \\
                Metric & Multi-Log Loss \\
                \# Classes & 2 \\
                Metric Frequency & 1 \\
                \# Leaves & 10 \\
                Max Depth & 10 \\
                Min Child Samples & 10 \\
                Learning Rate & 0.1 \\
                Feature Fraction & 0.9 \\
                Bagging Fraction & 0.8 \\
                Bagging Frequency & 5 \\
                \bottomrule
            \end{tabular}
        }
    \end{minipage}
\end{table*}

A classifier can be used to assess the similarity between two event categories from AHCAL with many correlated observable properties. The classifier's performance can be measured using the receiver-operating curve (ROC) and its area-under-curve (AUC). The ROC measures the true positive rate against the false positive rate, while the AUC indicates classifier performance. An AUC of 0.5 signifies random guessing, and a classifier with an AUC greater than 0.5 is more effective. The AUC can, therefore, be used to quantify the overall difference between a 'real' and a 'synthetic' neutral. 

 
A study was conducted using a training sample consisting of \qty{40}{\giga \electronvolt} $\pi^{-}$ and the entire $K^{0}_{L}$ sample from simulation, as shown in Table \ref{tab:EventTable}. This dataset was divided into training, validation, and test samples, with event details listed in Table \ref{tab:EventTable_BDT}. The study utilised the standard CALICE AHCAL Particle Identification (PID) classifier, which employs a gradient-boosted decision tree implemented in the \texttt{LightGBM} framework \cite{CALICESoft, PID, lightGBM}. This classifier accurately categorises hadrons, electrons, and muon-like events observed with AHCAL using thirteen event-level variables: $E_{\text{sum}}$;
The total number of hits in the event; the average hit radius; $\text{CoG}_{K}$; the fraction of $E_{\text{sum}}$ deposited in the first 22 AHCAL layers;  $K_{S}$; the fraction of $E_{\text{sum}}$ deposited after $K_{S}$; the fraction of $E_{\text{sum}}$ in the 'core' of a hadron shower ($R_{\mathrm{hit}}<1$ cell, $\geq 2$ adjacent hits, and $>0$ cells active in the same layer); the 'track-like' fraction of $E_{\text{sum}}$ ($\geq 2$ adjacent hits and 0 cells active in the same layer); the 'detached' fraction of $E_{\text{sum}}$: (0 adjacent hits); the number of hits in the event occurring after $K_{S}$; the number of hits contributing to the 'track-like fraction' of $E_{\mathrm{sum}}$ and number of hits contributing to the event in the last 4 AHCAL layers. 

\begin{figure*}

    \centering
    \includegraphics[width=0.49\linewidth]{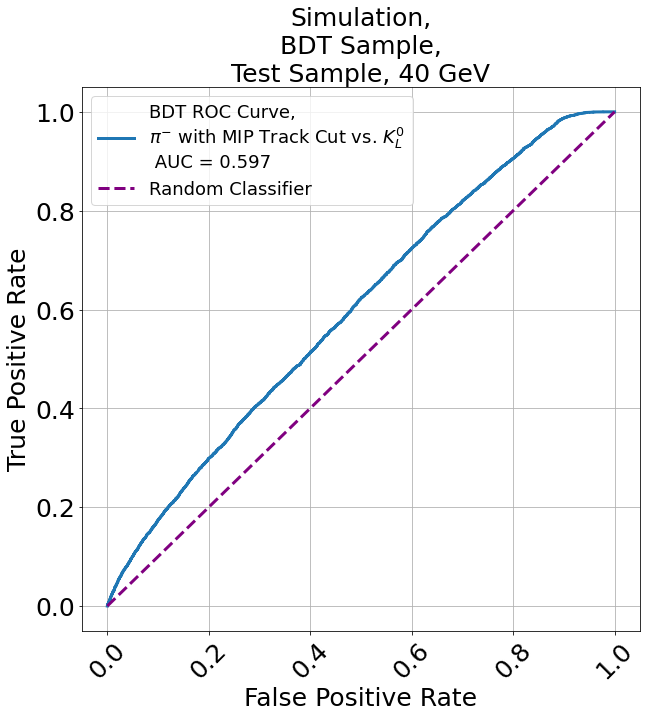}
    \caption{ ROCs from the trained classifier applied to the test sample from Table \ref{tab:EventTable_BDT}. The blue line indicates the performance of the model applied to the same testing sample of $K^{0}_{L}$ and $\pi^{-}$ with the MIP-track cut applied to the event variables specified. The purple dashed line indicates the expected curve for a random classifier.}

    \label{fig:BDTROC} 
\end{figure*}

The gradient-boosted decision tree classifier was re-optimised for classifying simulated $K^{0}_{L}$ from $\pi^{-}$ hadron showers with the MIP cut applied in AHCAL using the hyperparameters and loss detailed in Table \ref{tab:BDTHyperparams}. The re-optimisation was performed until the loss no longer improved after 5 training steps. Then, the newly optimised classifier was evaluated using the test dataset of $\pi^{-}$ hadron showers with the MIP cut applied. The classifier's performance was evaluated using the AUC.

The results are shown in Fig.~\ref{fig:BDTROC}. They illustrate that the classifier achieved an AUC of 0.597 for classifying $\pi^{-}$ with the MIP cut applied and $K^{0}_{L}$ showers in the test sample of Table \ref{tab:EventTable_BDT}.  It may be concluded that while these showers have some differences, they are challenging to distinguish at the event level. This result indicates the cut can produce 'convincing' synthetic neutrals.

\newpage

\subsection{Producing Events with a Charged and Synthetic Neutral Shower}\label{sec:ShowerAlg}

A method to combine a sample of single $\pi^{-}$ hadron showers into a shower with a charged hadron and a synthetic neutral hadron shower is presented.

 Firstly, two $\pi^{-}$ hadron showers are selected by a weighted random subsample of either the training, validation or testing sample of Table \ref{tab:EventTable} to produce the corresponding sample for training, validating and testing the neural networks. One is designated the charged candidate, $Q$; the other the synthetic neutral candidate, $N$. The weight for each particle energy is selected such that approximately equal numbers of each possible combination of particle energies appear in the final sample. Next,  the MIP-track cut is applied to $N$ only.

\begin{figure*}
    \centering
    \subfloat[
      \label{fig:dEdS}
    ]{\includegraphics[width=0.49\linewidth]{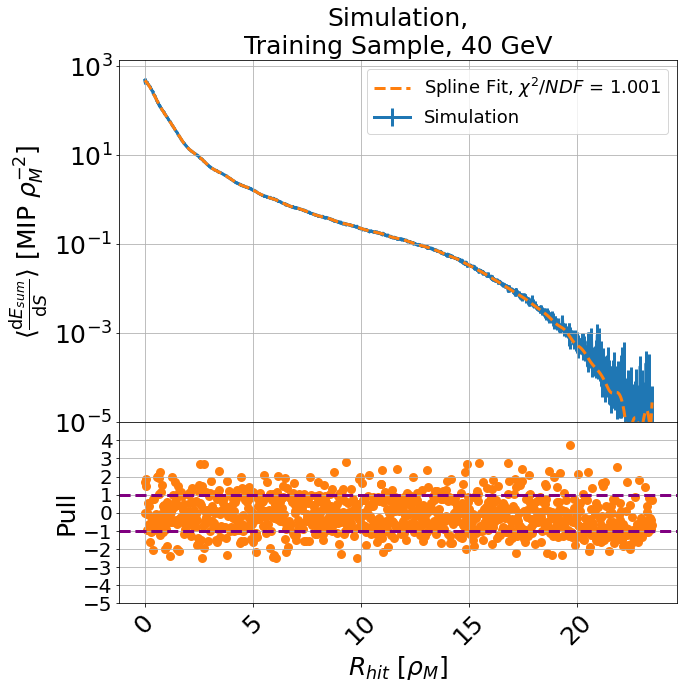}}
    \hfill
    \subfloat[
    \label{fig:RSep}
    ]{\includegraphics[width=0.49\linewidth]{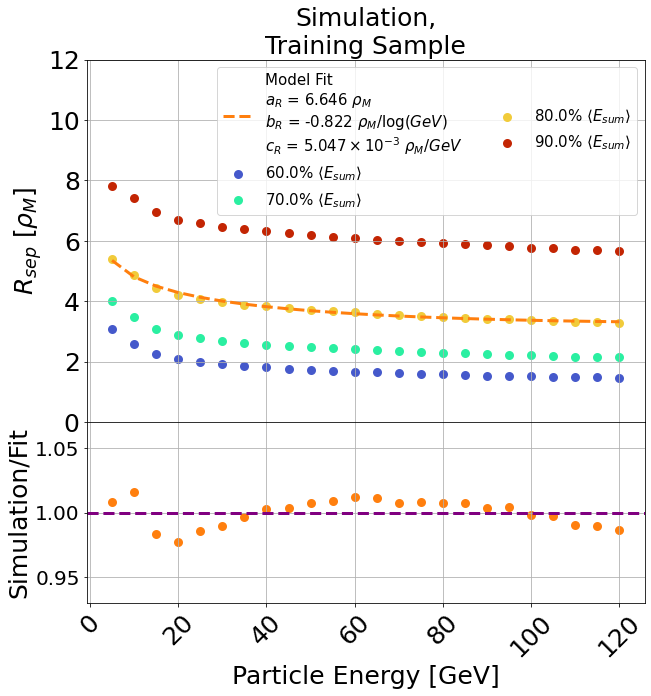}}
     \caption{Fig.~\ref{fig:dEdS} shows the differential energy loss per unit surface area as a function of $R_{\mathrm{hit}}$ for the \qty{40}{\giga \electronvolt} training sample of $\pi^{-}$ simulation in Table \ref{tab:EventTable}. The blue points show the simulation, and the orange dashed line indicates a spline fit to the points. The residues of the fit are shown in the bottom subplot. Fig.~\ref{fig:RSep} shows the value of $R_{\mathrm{sep}}$ as a function of particle energy for the simulation sample of Table \ref{tab:EventTable}. The blue, teal, orange, and red circle markers indicate the \qty{60}{\percent},  \qty{70}{\percent}, \qty{80}{\percent}, and \qty{90}{\percent} percentiles of $\langle E_{\mathrm{sum}} \rangle$. The dashed orange line indicates an ad-hoc fit to the \qty{80}{\percent} percentile.}
\end{figure*}

 Four integers, $\Delta I^{Q}$, $\Delta I^{N}$, $\Delta J^{Q}$ and $\Delta J^{N}$, are then defined as distances in cells by which to displace both $Q$ and $N$ in the $I$ and $J$ directions in calorimeter space to mimic the particles entering at a different position than the beam spot which is centred around $I = J = \qty{12.5}{\ensuremath{\text{cells}}}$ for both simulation and data. These displacements are first uniformly sampled within a circle with a radius $R_{\mathrm{circ}}$ centred at $I_{\mathrm{hit}}=\qty{0}{\text{cell}}$, $J_{\mathrm{hit}}=\qty{0}{\text{cell}}$, within the circumference of a circle of radius $R_{\mathrm{circ}}$.  The radius was chosen so that the showers have a moderate amount of confusion during training on average but not too much so that the training becomes overly challenging.

\begin{figure*}
    \centering
    \includegraphics[width=\linewidth]{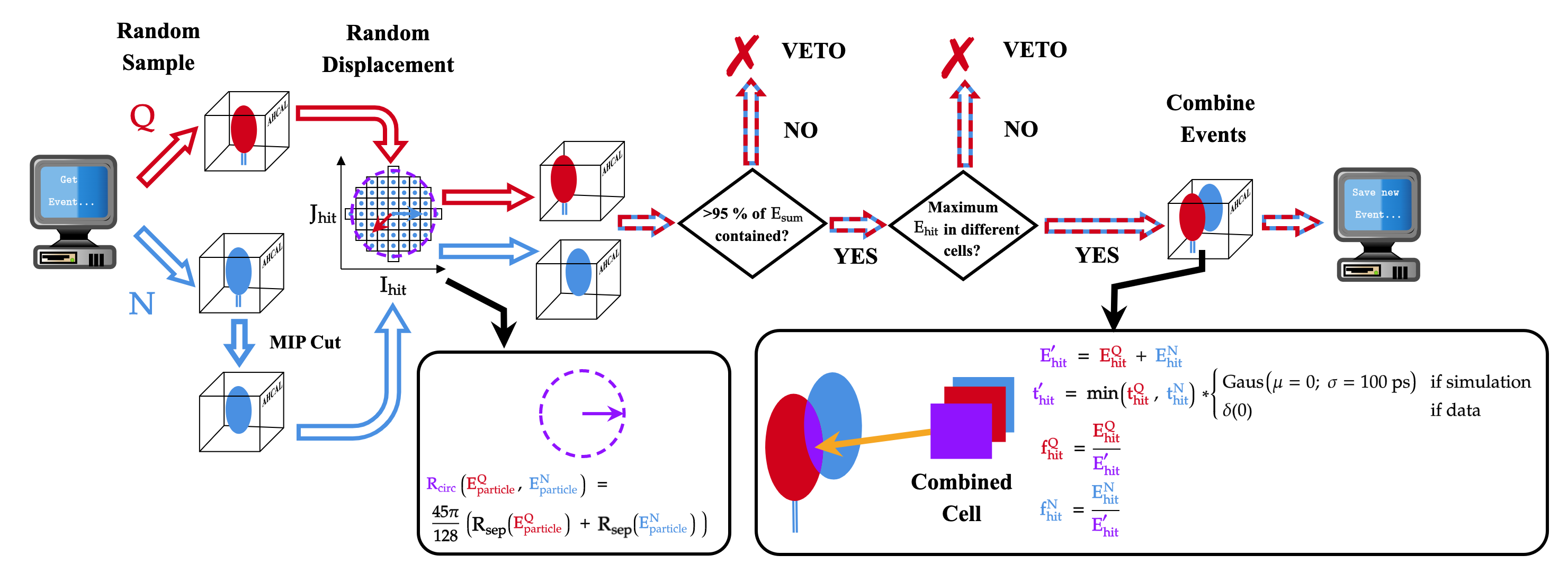}
    \caption{Diagram illustrating the shower-combination algorithm used to produce synthetic charged-neutral hadron shower events from a sample of single $\pi^{-}$ hadron showers observed with AHCAL. The red and blue arrows indicate the paths the charged and synthetic neutral hadron shower took. The squared circle indicates the distribution of available cells by which the event can be displaced from its entry position.  }
    \label{fig:Algorithm}
\end{figure*}

 The average distance between two points uniformly sampled within a circle of radius $R_{\mathrm{circ}}$ is denoted as $R_{\mathrm{sep}}$ for each shower. The method of estimating $R_{\mathrm{sep}}$ for each particle energy in the simulation training sample of Table \ref{tab:EventTable} is presented as follows. First, the differential energy loss by a hadron shower per unit area of a circle (i.e., a thin ring) around the centre-of-gravity is calculated, defined as $\left\langle \mathrm{d}E_{\mathrm{sum}}/\mathrm{d}S \right\rangle$, where $\mathrm{d}S = 2 \pi R_{\mathrm{hit}}\,\mathrm{d}R_{\mathrm{hit}}$ with $\mathrm{d}R_{\mathrm{hit}}$ as a bin width determined using the Freedman-Diaconis bin rule \cite{Binning}. An example is shown in Fig.~\ref{fig:dEdS}. A spline is fitted to the distribution, and then cumulatively integrated as a function of radial distance from the shower's centre-of-gravity, $\langle E_{\mathrm{sum}} (R_{\mathrm{hit}}) \rangle = \int^{R_{\mathrm{hit}}}_{0} \left\langle \mathrm{d}E_{\mathrm{sum}}/\mathrm{d}S \right\rangle \cdot 2\pi R_{\mathrm{hit}} \,\mathrm{d}R_{\mathrm{hit}}$. The $R_{\mathrm{sep}}$ value for a particular $E_{\mathrm{particle}}$ is determined where this cumulative integral reaches \qty{80}{\percent} of $\left \langle E_{\mathrm{sum}} \right \rangle$.
 

The relationship between the particle energy and $R_{\mathrm{sep}}$ is expected to decrease with particle energy as the electromagnetic fraction of a hadron shower increases, making the shower more energy-dense. This relationship can be approximated using an ad-hoc function, $R_{\mathrm{sep}}(E_{\mathrm{particle}}) = a_{R} + b_{R}\cdot \log{E_{\mathrm{particle}}} + c_{R} \cdot E_{\mathrm{particle}}$, where $a_{R}$, $b_{R}$, and $c_{R}$ are free parameters obtained from a fit. The relationship for simulation is shown in Fig.~\ref{fig:RSep}.

Finally, $R_{\mathrm{circ}}$ is calculated using the formula $R_{\mathrm{circ}}(E_{Q}, E_{N}) = 45\pi/128 \cdot (R_{\mathrm{sep}}(E_{Q}) + R_{\mathrm{sep}}(E_{N})) $, where the factor relates a circle's radius to the average distance between two uniformly sampled points within its circumference \cite{45pi}. The vectors $\{\Delta I^{Q}$, $\Delta J^{Q} \}$ and  $\{\Delta I^{N}$, $\Delta J^{N} \}$ obtained by rejection sampling using $\Delta I^{2} + \Delta J^{2} \leq R^{2}_{\mathrm{circ}}$.

Next, all of the hits and track positions of $Q$ and $N$ are then shifted by the two sampled integers. For example, all the hits in $Q$ are displaced by $\Delta I^{Q}$ and $\Delta J^{Q}$, (i.e. $I^{Q}_{\mathrm{hit}} \rightarrow I^{Q}_{\mathrm{hit}} + \Delta I^{Q}$,  $J^{Q}_{\mathrm{hit}} \rightarrow J^{Q}_{\mathrm{hit}} + \Delta J^{Q}$), and the track position of the charged particle is also shifted by the same integers (i.e. $I^{Q}_{\mathrm{track}} \rightarrow I^{Q}_{\mathrm{track}} + \Delta I^{Q}$,  $J^{Q}_{\mathrm{track}} \rightarrow J^{Q}_{\mathrm{track}} + \Delta J^{Q}$).

\begin{figure*}
    \centering
    \subfloat[
      \label{fig:RSep_Dist}
    ]{\includegraphics[width=0.49\linewidth]{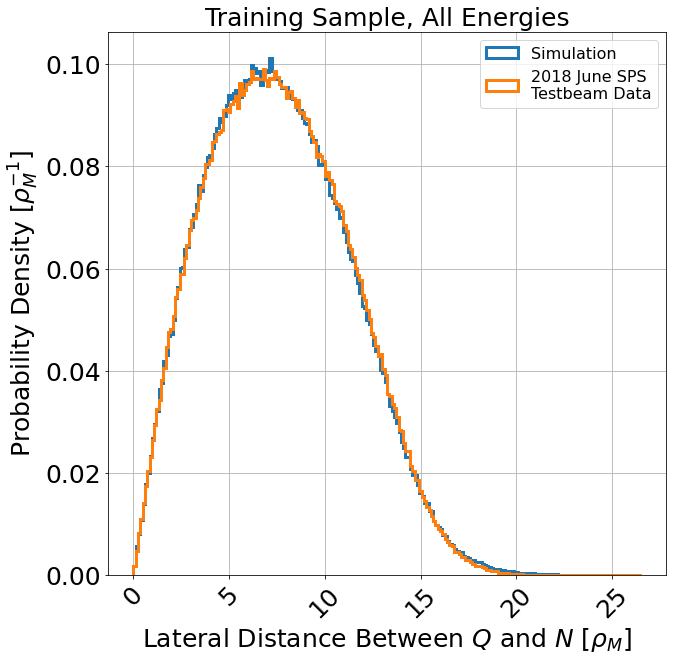}}
    \hfill
    \subfloat[
    \label{fig:RSep_CoG_Sim}
    ]{\includegraphics[width=0.4\linewidth]{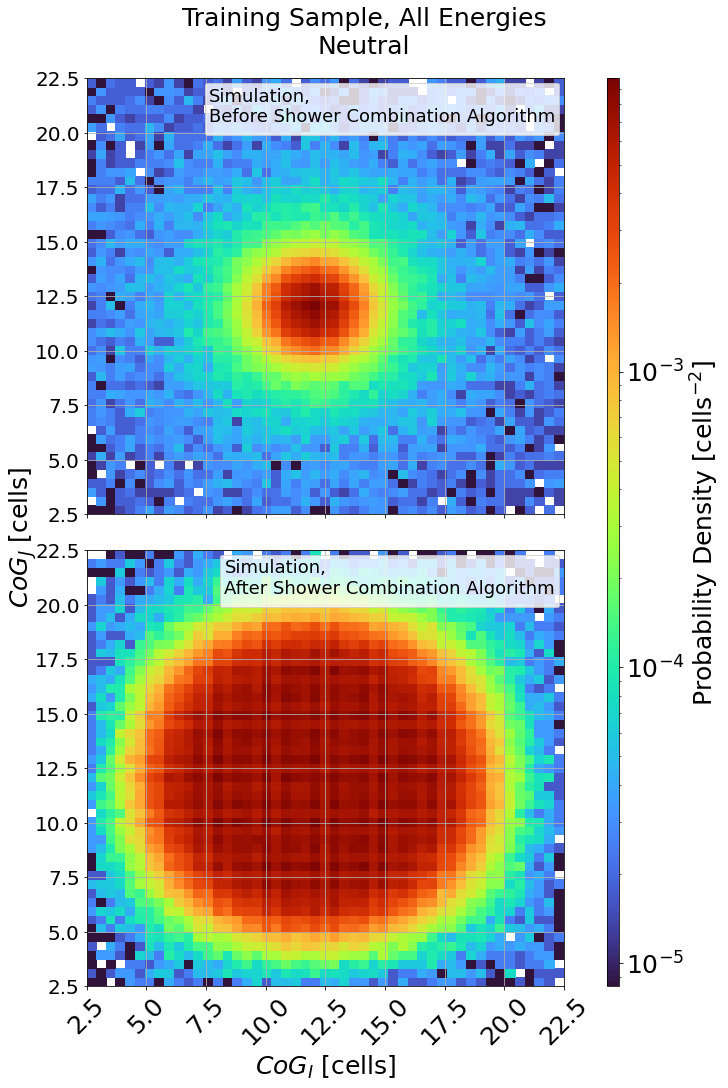}}
    \subfloat[
    \label{fig:RSep_CoG_Data}
    ]{\includegraphics[width=0.4\linewidth]{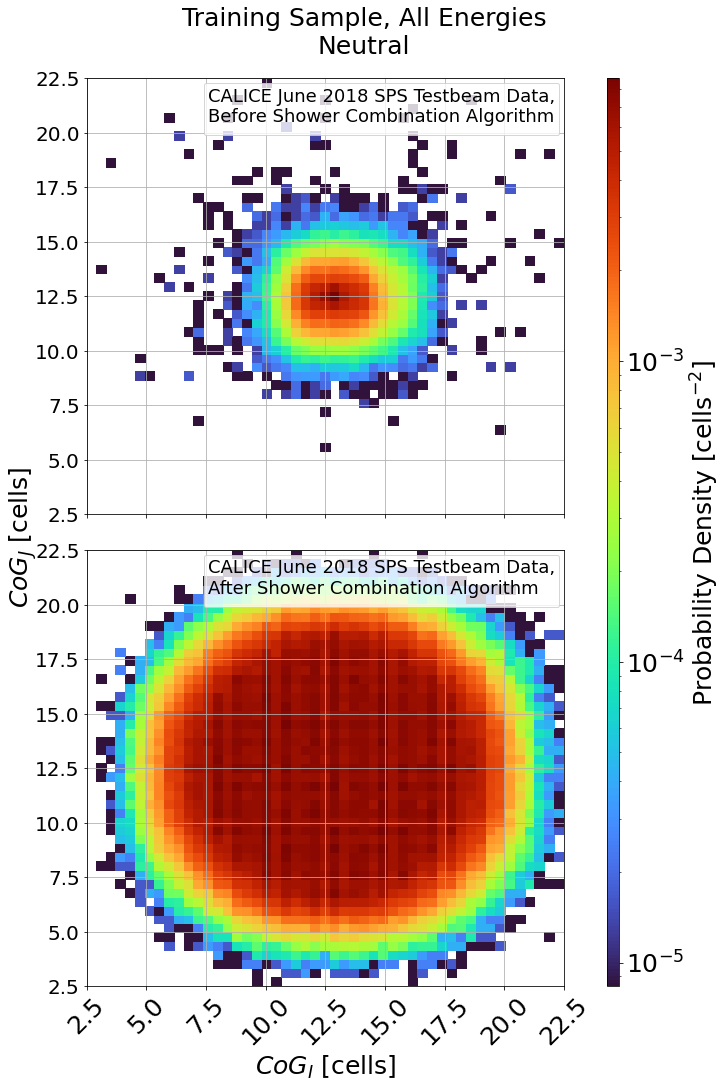}}
     \caption{Fig.~\ref{fig:RSep_Dist} shows the distribution of distances between $Q$ and $N$ in Moliere radii ($\rho_{\mathrm{M}}$). The blue and orange lines indicate simulation and data, respectively. Figs.~\ref{fig:RSep_CoG_Sim} and \ref{fig:RSep_CoG_Data} show the distribution of the centres-of-gravity of the synthetic neutral shower before (top plot) and after (bottom plot) the shower overlay algorithm is applied, for simulation and data respectively. The colour axis indicates probability density. }
     \label{fig:ShowerAlgExamples}
\end{figure*}

At this stage, any hits outside the calorimeter are removed from both showers. If over 95\% of each shower's energy remains in the calorimeter and if the highest energy cells are not shared between $Q$ and $N$, the algorithm continues. Event-level properties like the centre-of-gravity are recalculated based on the MIP-track cut and the removed hits. If the criteria are unmet, the event is rejected, and displacement integers are resampled until they meet the criteria.

Upon satisfying the criteria, $Q$ and $N$ are combined. The energies from $Q$ and $N$ are added for cells with shared energy. The minimum hit time of $Q$ and $N$ is taken for the cell, with a random Gaussian smearing of \qty{100}{\pico \second} applied after selection in the simulation. As the hit time in data already includes the detector's time resolution, no additional smearing is applied. Finally, energy fractions for each hadron shower, $f^{Q}_{\mathrm{hit}}$ and $f^{N}_{\mathrm{hit}}$, are calculated from the combined energy. The result is an event containing a charged and synthetic neutral hadron shower from a dataset of single $\pi^{-}$ hadron showers, useful for training and validating machine learning algorithms for shower separation. A diagram of the algorithm is shown in Fig.~\ref{fig:Algorithm}.

Fig.~\ref{fig:RSep_Dist} shows that a wide range of inter-shower distances are available in the training dataset. The most probable distance between $Q$ and $N$ is $\qty{5.5}{\ensuremath{\rho_{\mathrm{M}}}}$ for both simulation (left) and data (right), determined by kernel density estimate. Figs.~\ref{fig:RSep_CoG_Sim}-\ref{fig:RSep_CoG_Data} show that the distribution of the initial centres-of-gravity of the hadron showers in the calorimeter is convolved with a circle function, indicating a wide variety of shower configurations in the final training sample.

\clearpage

\section*{Acknowledgments}

 We would like to thank the technicians and the engineers who contributed to the design and construction of the CALICE AHCAL prototype detector. We also gratefully acknowledge the CERN management for its support and hospitality and its accelerator staff for the reliable and efficient operation of the test beam. The authors acknowledge the support from the BMBF via the High-D consortium. This work is supported by the Deutsche Forschungsgemeinschaft (DFG, German Research Foundation) under Germany's Excellence Strategy, EXC 2121, Quantum Universe (390833306).





\end{document}